\documentclass[aps,pra,twocolumn,amsmath,showpacs]{revtex4}
\usepackage{graphicx}
\begin{document}
\draft
\title{Scattering properties of weakly bound dimers of fermionic atoms} 
\author{D.S. Petrov$^{1,2,3}$, C. Salomon$^4$, and G.V. Shlyapnikov$^{2,3,5,6}$}
\affiliation{$^1$ITAMP, Harvard-Smithsonian Center for Astrophysics, and Harvard-MIT
Center for Ultracold Atoms, \\
Department of Physics, Harvard University, Cambridge, MA 02138, USA\\
$^2$Kavli Institute for Theoretical Physics, University of California, Santa Barbara,
CA 93106-4030, USA\\
$^3$Russian Research Center Kurchatov Institute, Kurchatov Square, 123182 Moscow,
Russia\\
$^4$Laboratoire Kastler-Brossel, Ecole Normale Sup\'erieure, 24 rue Lhomond. 75005
Paris, France\\
$^5$\mbox{Laboratoire Physique Th\'eorique et Mod\`eles Statistique, Universit\'e
Paris Sud, B\^at. 100, 91405 Orsay, France}\\
$^6$\mbox{Van der Waals-Zeeman Institute, University of Amsterdam, Valckenierstraat
65/67, 1018 XE
Amsterdam, The Netherlands}}

\date{\today}
\begin{abstract}
We consider weakly bound diatomic molecules (dimers) formed in a two-component atomic
Fermi gas with a large positive scattering length for the interspecies interaction.
We develop a theoretical approach for calculating atom-dimer and dimer-dimer elastic
scattering and for analyzing the inelastic collisional relaxation of the molecules
into deep bound states. This approach is based on the single-channel zero range
approximation, and we find that it is applicable in the vicinity of a wide two-body
Feshbach resonance. Our results draw prospects for various interesting manipulations
of weakly bound dimers of fermionic atoms.    
\end{abstract}

\maketitle

\section{Introduction} \label{sec.int}

The studies of degenerate atomic Fermi gases attract a great deal of interest as they
are directed to reveal novel macroscopic quantum states and provide links between
quantum gases and condensed matter systems. Experiments with ultracold two-component
clouds of fermionic atoms  widely use Feshbach resonances for the intercomponent
interaction (scattering length $a$). This allows one to switch the sign and tune the
absolute value of $a$ which at resonance changes from $+\infty$ to $-\infty$. On the
positive side of the resonance ($a>0$), one expects the formation of weakly bound
diatomic molecules, which represent composite bosons and can form a Bose-Einstein
condensate (BEC). On the negative side ($a<0$), the  interaction between atoms of
different components is attractive and they should undergo the well-known
Bardeen-Cooper-Schrieffer (BCS) superfluid pairing at sufficiently low temperatures.
The BCS transition temperature is much lower than the Fermi energy, which makes it
difficult to achieve this transition. 

Molecular BEC for $a>0$ and atomic BCS pairing for $a<0$, describe the system
sufficiently far from the resonance. In the vicinity of the resonance, where the
density and the scattering length satisfy the inequality $n|a|^3\agt 1$, the gas
enters a strongly interacting regime. This BCS-BEC cross-over regime has been
discussed in the literature in the context of superconductivity
\cite{Eag,Leg,Noz,Rand} and for superfluidity in two-dimensional $^3$He films
\cite{M,MYu}. Atomic Fermi gases in the strongly interacting regime are expected to
have a  comparatively high superfluid transition temperature
\cite{Holland1,Tim,Holland2,Holland3,Griffin} and are characterized by a universal
behavior of interactions \cite{Heis,Carlson,duke1,Bourdel} and a universal
thermodynamics \cite{Ho1}. They are now actively being studied in relation to the
nature of superfluid pairing \cite{Falco,Bruun,Bulgac,Perali,Chen,Holland4,Ho2}. 

Recent investigations of two-component $^{40}$K and $^6$Li atomic Fermi gases are
marked by remarkable achievements. Those include the formation
\cite{ens1,rudy1,randy1,jila1} and Bose-Einstein condensation
\cite{jila2,rudy2,mit1,ens2,randy2} of long-lived weakly bound diatomic molecules
(dimers) on the positive side of a Feshbach resonance  ($a>0$), and a BEC-type
behavior of fermionic atom pairs in the strongly interacting regime
\cite{jila3,mit2}.  An anomalous dependence of frequencies and damping rates of
quadrupole excitations on the interaction strength, observed in experiments
\cite{duke2,rudy3}, may be a signature for a transition from a superfluid to
collisionless regime. Strong evidence for the superfluid regime was obtained in the
Innsbruck experiment \cite{rudy4} through the radio-frequency measurement of the
pairing gap for a strongly interacting Fermi gas of $^6$Li. 

The studies of the strongly interacting regime for the BCS-BEC cross-over require the
knowledge of many-body correlations. In particular, one should reproduce a correct
equation of state in the limit of BEC of a weakly interacting gas of dimers for
$a>0$ (see \cite{Holland4}) . Hence, one should know the interaction
between these dimers. For a large $a$ they are weakly bound and have a  large size
($\sim a$) which greatly exceeds the characteristic radius  of interaction between
the atoms.  In our previous work \cite{Petrov1} we have outlined a  method for
studying the elastic interaction between such molecules and their collisional
relaxation to deep bound states. This method is based on the zero-range
approximation, and the dimer-dimer scattering length is found to be $0.6a$. The
imaginary part of the scattering amplitude, originating from the collisional
relaxation, is extremely small.  Being in the highest ro-vibrational state these
diatomic molecules are characterized by a remarkable collisional stability. The
physical reason is the Pauli principle in combination with a large size of the
molecular state (small momenta of bound fermionic atoms): collisional relaxation is
suppressed as it requires at least two identical fermions with small momenta to
approach each other to a short distance \cite{Petrov1}. 

In this paper we present a detailed analysis of elastic and inelastic atom-dimer and
dimer-dimer interactions in the zero-range approximation. The paper is organized as
follows. In Section II we introduce the zero-range approximation in a way it has
been done for the  two-body problem (see, for example, \cite{Bethe,Huang}). Section
III contains a general description of the zero-range approximation for the case of
three particles. In Section IV we review the problem of atom-dimer elastic
scattering in the zero-range approximation, and in Section V present a derivation
for the relaxation of the dimers to deep bound states in atom-dimer collisions.
Sections VI and VII contain a generalization of the zero-range approximation to the
case of four particles. In these Sections we present a detailed derivation of the
results for the elastic dimer-dimer interaction and for the relaxation of the dimers
to deep bound states in dimer-dimer collisions. In Section VIII we show that our
results can be used for weakly bound diatomic molecules obtained in two-component
atomic Fermi gases by using wide Feshbach resonances, and in Section IX we conclude.

\section{Zero-range approximation for the two-body problem} \label{sec.twobody}

We first follow well known results and introduce the zero-range approximation as this
has been done for the two-body problem (see \cite{Bethe,Huang}). We consider elastic
pair collisions between cold distinguishable atoms interacting with each other 
via a spherically symmetric potential and assume that their de Broglie
wavelength is much larger than the characteristic radius of this potential,
$R_e$. In other words, we have the condition $kR_e\ll 1$, where $k$ is the relative
wavevector of the atoms. In this case the scattering is dominated by the $s$-wave
contribution. The behavior of the wavefunction at interatomic distances $r\gg R_e$
is governed by the scattering length $a$ which is related to the scattering phase
shift as $\delta=-\arctan(ka)$. Given the scattering length, the details of the
interatomic potential at distances $r\alt R_e$ are practically irrelevant for
scattering parameters and give rise to corrections of the order of $kR_e$ or
smaller for the scattering amplitude \cite{LL3}. 

The key idea of the zero-range approximation is to solve the equation for the free
relative motion of two atoms placing the Bethe-Peierls boundary condition on the
wavefunction at a vanishing  $r$:
\begin{equation}\label{twobody.Bethe-Reierls}
\frac{(r\psi)'}{r\psi}=-\frac{1}{a},\,\,\,\,\,\,\,r\rightarrow 0,
\end{equation}
which can also be rewritten as
\begin{equation}\label{twobody.boundary1}
\psi\propto (1/r-1/a),\,\,\,\,\,\,\,r\rightarrow 0.
\end{equation}
One then gets a correct expression for the wavefunction at distances $r\gg R_e$.

For the free relative motion of two colliding atoms, the Schr\"odinger equation takes
the form
\begin{equation}  \label{schrfree}
-\left(\nabla_{{\bf r}}^2+k^2\right)\psi=0.
\end{equation}
A general solution of this equation for our scattering problem, which is valid at any 
finite $r$, is given by
\begin{equation}  \label{greensol}
\psi=\exp(i{\bf kr})+hG({\bf r},0),
\end{equation}
where $G({\bf r},{\bf r}')=(1/4\pi|{\bf r}-{\bf r}|')\exp(ik|{\bf r}-{\bf r}'|)$ is
the Green function representing the solution of Eq.~(\ref{schrfree}) with the right
hand side equal to $\delta({\bf r}-{\bf r}')$. The $s$-wave part of the incident
wave given by the first term on the right hand side of Eq.~(\ref{greensol}), is equal
to
$\sin{kr}/kr$. As the wavefunction $\psi$ should satisfy the boundary condition
(\ref{twobody.Bethe-Reierls}) at $r\rightarrow 0$, for the coefficient $h$ we
immediately obtain
\begin{equation} \label{h}
h=-\frac{a}{1+ika}=\frac{\exp(2i\delta)-1}{2ik}.
\end{equation}   
Then Eq.~(\ref{greensol}) reproduces the well-known result for the $s$-wave part of
the wavefunction at $r\gg R_e$:
\begin{equation} \label{psikr} 
\psi\propto \frac{\sin(kr+\delta)}{kr}.
\end{equation}
The use  of the zero-range approximation is especially important for the case of
resonance scattering characterized by the scattering length $|a|\gg R_e$. Then, for
interparticle distances in the interval 
$R_e\ll r\ll 1/k$, from Eq.~(\ref{psikr}) one finds that
Eq.~(\ref{twobody.boundary1}) gives a correct result for $\psi$ at distances of the
order of or even much smaller than $|a|$.

For a large positive scattering length $a\gg R_e$, there is a weakly bound state of
two atoms. The binding energy $\varepsilon_0$ and wavefunction $\phi_0$ of this
state  at distances  $r\gg R_e$ can also be found in the zero-range approximation. 

The free relative motion of atoms in the bound state is described by the
Schr\"odinger equation:
\begin{equation}  \label{schrbound}
(-\nabla_{{\bf r}}^2+m\varepsilon_0/\hbar^2)\phi_0=0,
\end{equation} 
where $m$ is the atom mass, and the (negative) energy of the molecular state is
$E=-\varepsilon_0$.  A general solution of Eq.~(\ref{schrbound}) at any finite $r$
can be written as
\begin{equation} \label{genbound}
\phi_0(r)=\psi_0({\bf r})+h_0G_{\varepsilon_0}({\bf r},0),
\end{equation}
where $G_{\varepsilon_0}({\bf r},{\bf r}')=(1/4\pi|{\bf r}-{\bf
r}'|)\exp(-\sqrt{m\varepsilon_0/\hbar^2}|{\bf r}-{\bf r}'|)$ is the Green function
of Eq.~(\ref{schrbound}), and $\psi_0$ is the solution of this equation that is
finite and regular at any distance including $r=0$. For $r\rightarrow 0$, the
wavefunction (\ref{genbound}) should satisfy the boundary condition
(\ref{twobody.Bethe-Reierls}).

One can easily see that any non-trivial solution of Eq.~(\ref{schrbound}), finite at
$r\rightarrow\infty$, behaves as $1/r$ for $r\rightarrow 0$. Therefore, we have
$\psi_0=0$. Then, using the boundary condition (\ref{twobody.Bethe-Reierls})
 for $\phi_0(r)$ at vanishing $r$, one finds that the binding energy of the weakly
bound state is
\begin{equation}\label{twobody.benergy}
\varepsilon_0=\hbar^2/ma^2,
\end{equation} 
and the wavefunction $\phi_0$ normalized to unity is given by
\begin{equation}\label{twobody.boundstate}
\phi_0(r)=(1/\sqrt{2\pi a}r)\exp(-r/a).
\end{equation}
Note that under the condition $|a|\gg R_e$ the main contribution to the normalization
integral comes from distances $r\gg R_e$, where Eq.~(\ref{twobody.boundstate}) is
valid. Relative corrections to the binding energy $\varepsilon_0$ are of the order
of $(R_e/a)$.

\section{General formalism for three fermions} \label{sec.general} 

Theoretical studies of the 3-body problem have a long prehistory (see \cite{fedorov}
for a review).
In this Section we consider a three-body system consisting of two identical fermions
interacting
with a third one, which is not identical to the first two, via a short-range pair
isotropic potential $U(r)$. The fermions have the same mass and we will denote
the identical ones by the symbol $\uparrow$ and the third one by the symbol
$\downarrow$. In
the center of mass reference frame the state of the system with total energy $E$ is
described by the Schr\"odinger equation
\begin{equation}\label{general.schrodinger}
\left[-\nabla_{\bf
x}^2-\nabla_{\bf y}^2+\sum_\pm U\left(\frac{\sqrt{3}\,{\bf x}\pm{\bf
y}}{2}\right)-E\right]\Psi=0,
\end{equation}
where ${\bf y}$ is the distance between the identical fermions, and $\sqrt{3}{\bf
x}/2$
is the distance between their center of mass and the third atom. Hereinafter we use
notations in which $\hbar=m=1$. 

The wavefunction $\Psi$ is antisymmetric with respect to the permutation of identical
fermions, i.e. it changes sign under the transformation $ {\bf y}\rightarrow -{\bf
y}$.

We will discuss the case of resonant two-body interaction, that is, we assume that
the two-body problem for the interaction potential $U(r)$ is characterized by a
large scattering length 
\begin{equation}  \label{aRe}
|a|\gg R_e.
\end{equation} 
As has been shown by Efimov \cite{efimov}, in this case short-range physics is not
important and the
3-body problem is universal in the sense that it can be described in terms of the
2-body scattering length. One can then use the zero range approximation for the
interatomic potential. This was first done even earlier by Skorniakov and
Ter-Martirosian \cite{STM} in relation to the neutron-deutron scattering, which is
similar to elastic scattering of atoms by weakly bound dimers. The work of Ref.
\cite{STM} was followed by related discussions \cite{Danilov}. In this Section, we
present the form of the zero range approximation for 3-body systems, that was
outlined in Refs. \cite{Petrov2,Petrov1} for the 3-body recombination and atom-dimer
scattering. 

Under the condition (\ref{aRe}), the zero range approximation is applicable even at
interparticle distances much smaller than $|a|$, as long as these distances greatly
exceed $R_e$. Also, this approximation properly describes weakly bound states of two
particles at $a>0$.  
According to the zero-range approximation, Eq.~(\ref{general.schrodinger}) is
equivalent to the Poisson equation
\begin{equation}\label{general.poisson}
-[\nabla_{\bf x}^2+\nabla_{\bf y}^2+E]\Psi=0,
\end{equation}
with the boundary condition (\ref{twobody.boundary1}) set for a vanishing distance
between any of the two distinguishable fermions, i.e. for ${\bf
r}_{\pm}=(\sqrt{3}\,{\bf x}\pm{\bf
y})/2\rightarrow 0$. Taking into account the symmetry we can write the boundary
conditions at the two boundaries as
\begin{equation}\label{general.boundary}
\Psi\approx \pm\frac{1}{4\pi}f\left( {\bf
r}\right)\left(\frac{1}{r_{\pm}}-\frac{1}{a}\right),\,\,\,\,\,\,r_{\pm}\rightarrow
0,
\end{equation}
where ${\bf r}=\mp2{\bf y}/\sqrt{3}$ for $r_{\pm}\rightarrow 0$. The function $f$
contains information about the relative motion of a
$\uparrow$-fermion with respect to the two other atoms when they are on top of each
other. In the case of atom-dimer scattering the function $f$ plays a role of the
wavefunction of the atom-dimer  relative motion. 

The Green function of Eq.~(\ref{general.poisson}), that is the solution of this
equation with the right hand side equal to $\delta({\bf x}-{\bf x}')\delta({\bf
y}-{\bf y}')$,
is given by
\renewcommand{\arraystretch}{2} 
\begin{equation}\label{general.green}
G_E(X)=\left\{\begin{array}{lr}-(8\pi^3X^2)^{-1}EK_2(\sqrt{-E}X),&E<0\\
i(16\pi^2X^2)^{-1}EH_2(\sqrt{E}X).&E>0\end{array}\right.
\end{equation}
\renewcommand{\arraystretch}{1}
\noindent Here  $X=\sqrt{({\bf x}-{\bf x}')^2+({\bf y}-{\bf y}')^2}$, $K_2$ is an
exponentially decaying
Bessel function, and $H_2$ is a Hankel function representing an outgoing wave. For
$\sqrt{|E|}X\ll 1$ we have
\begin{equation} \label{G0}
G_0=(4\pi^3X^4)^{-1}.
\end{equation}

As in the two-body case described by Eqs.~(\ref{greensol}) and (\ref{genbound}), a
general solution of Eq.~(\ref{general.poisson}) at finite distances between
$\uparrow$ and $\downarrow$ fermions can be expressed through the Green function
$G_E (\ref{general.green})$, with coordinates ${\bf x}',{\bf y}'$ corresponding to a
vanishing distance between distinguishable fermions, that is for ${\bf
r}'_+\rightarrow 0$ and for ${\bf r}'_-\rightarrow 0$. We thus have
\begin{align}\label{general.psi}
&\Psi({\bf x},{\bf y})=\Psi_0({\bf x},{\bf y})+\nonumber \\
&\int G_E\Bigl(\!\sqrt{({\bf x}-{\bf r'}/2)^2+({\bf
y}+\sqrt{3}{\bf r'}/2)^2}\Bigr)h({\bf r'}){\rm d}^3r' -\nonumber \\
& \int G_E\Bigl(\!\sqrt{({\bf x}-{\bf r'}/2)^2+({\bf
y}-\sqrt{3}{\bf r'}/2)^2}\Bigr)h({\bf r'}){\rm d}^3r',
\end{align} 
where $\Psi_0$ is a properly symmetrized and finite solution of
Eq.~(\ref{general.poisson}), regular at vanishing distances between $\uparrow$ and
$\downarrow$ fermions. For a negative total energy $E$, non-trivial solutions of this
type do not
exist and we have to put $\Psi_0=0$. The function $h({\bf r})$ has to be determined
relying on the
boundary conditions (\ref{general.boundary}). 

For this purpose we consider the limit ${\bf r}_{+}\rightarrow 0$ and analyze the
leading behavior of the terms on the right hand side of Eq.~(\ref{general.psi}). The
argument of
the Green function in the third term can be
written as $\sqrt{r^2+r'^2+{\bf rr'}-\sqrt{3}{\bf r}_+{\bf r}'+r_+^2}$, where ${\bf
r}=({\bf x}-\sqrt{3}{\bf y})/2$, and this term is finite for $r_+\rightarrow
0$. It can be written as
\begin{equation} \label{term3}
\int G_E(\sqrt{r^2+r'^2-{\bf rr'}})h({\bf r}')d^3r'.
\end{equation} 
In the second term, the argument of the Green function takes the
form $\sqrt{({\bf r}-{\bf r}')^2+r_+^2}$. We then subtract from this term and add to
it an auxiliary quantity 
\begin{align} \label{termaux}
&h({\bf r})\int\! G_E\left(\sqrt{({\bf r}-{\bf r'})^2+r_+^2}\right){\rm d}^3r' 
\nonumber \\
&=(1/4\pi r_+)h({\bf r})\exp(-\sqrt{-E}r_+).
\end{align}
The result of the subtraction gives a quantity $I=\int G_E(\sqrt{({\bf r}-{\bf
r'})^2+r_+^2})[h({\bf r'})-h({\bf r})]d^3r'$.  For $r_+\rightarrow 0$,  this
quantity remains finite and can be written as 
\begin{equation} \label{termP}
I=P\int G_E(|{\bf r}-{\bf r'}|)[h({\bf r'})-h({\bf r})]d^3r', 
\end{equation}
where the symbol $P$ denotes the principal value for the integration over $dr'$. The
derivation of Eq.~(\ref{termP}) and the proof of the convergence of the integral on
the right hand side of this equation are given in  Appendix.    

Neglecting terms that are vanishing for $r_+\rightarrow 0$, the two last lines of
Eq.~(\ref{general.psi}) are  given by the sum of Eqs.~(\ref{term3}), (\ref{termaux}),
and (\ref{termP}). Thus, in the limit $r_+\rightarrow 0$ we can write
Eq.~(\ref{general.psi}) in the form containing a singular term proportional to
$r_+^{-1}$, and regular terms independent of $r_+$:
\begin{eqnarray}\label{general.boundary2}
&\Psi&\approx \frac{r_+^{-1}-\sqrt{-E}}{4\pi}h({\bf r})+\int\big\{G_E(|{\bf
r}-{\bf r'}|)[h({\bf r'})-h({\bf r})]\nonumber \\
&-&\, G_E(\!\sqrt{r^2+r'^2-{\bf r}{\bf r'}})h({\bf r'})\big\}{\rm d}^3r'+D({\bf
r})/4\pi,
\end{eqnarray}  
where $D({\bf r})=4\pi \Psi_0(-{\bf r}/2,\sqrt{3}{\bf r}/2)$, and
$\sqrt{-E}=-i\sqrt{E}$ for positive energies
$E$. Hereafter we omit the principal value symbol $P$ for the first term of the
integrand of Eq.~(\ref{general.boundary2}). 

As Eq.~(\ref{general.boundary2}) should coincide with Eq.~(\ref{general.boundary}) at
$r_+\rightarrow 0$, comparing the singular terms of these equations we  immediately
find that $h({\bf r})=f({\bf r})$. Matching the regular terms yields the
equation for the function $f$:
\begin{equation}\label{general.main}
(\hat L_E-a^{-1}+\sqrt{-E})f({\bf r})=D({\bf r}),
\end{equation}
where the integral operator $\hat L_E$ is given by
\begin{eqnarray}\label{general.L}
\hat L_E f({\bf r})&=&4\pi\int\big\{G_E(|{\bf r}-{\bf r'}|)[f({\bf r})-f({\bf
r'})]\nonumber \\
&+&\, G_E(\!\sqrt{r^2+r'^2-{\bf
r}{\bf r'}})f({\bf r'})\big\}{\rm d}^3r'.
\end{eqnarray} 

The operator on the left hand side of Eq.~(\ref{general.main}) conserves angular
momentum. Therefore, one can expand $f$ and $D$ in spherical harmonics and work only
with a
set of uncoupled equations for functions of a single variable $r$. The knowledge of
the function $f({\bf r})$ allows one to calculate all scattering parameters. In the
next section we demonstrate this by calculating the atom-dimer scattering length in
the ultracold limit.

\section{Atom-dimer elastic scattering} \label{sec.atommol}

According to the discussion in Section~\ref{sec.twobody}, for a large positive
scattering length $a\gg R_e$ one has a weakly bound molecular state of two
distinguishable fermions ($\uparrow$ and $\downarrow$), with the binding energy
$\varepsilon_0=1/a^2$ and the wavefunction $\phi_0$ described by
Eq.~(\ref{twobody.boundstate}). We now consider elastic collisions of these bosonic
dimers with $\uparrow$ (or $\downarrow$) fermionic atoms at collision energies
$\varepsilon<\varepsilon_0$. Then the total energy of the three-body atom-dimer
system is $E=(\varepsilon-\varepsilon_0)<0$ and collisions do not lead to
dissociation of the dimers.  

From Eq.~(\ref{twobody.boundstate}) one finds that the size of the weakly bound
molecular state is $\sim a$. For a large separation between the atom and the dimer,
the three-body wavefunction factorizes into a product:
\begin{equation}\label{atomdimer.asympt}
\Psi({\bf x},{\bf y})\approx \phi_0(r_+)\chi({\bf r}),\,\,\,\,\,\,\,r\gg a.
\end{equation}
A characteristic value of the distance $r_+$ between atoms in the dimer is $\sim a$,
and in the limit $r\gg a$ the atom-dimer separation is equal to $\sqrt{3}r/2$. The
wavefunction $\chi({\bf r})$ describes the relative atom-dimer motion and can be
represented as a superposition of an incident and scattered waves. For
$r\rightarrow\infty$ we have: 
\begin{equation}\label{atomdimer.f}
\chi({\bf r})\approx \exp(ikz)+\frac{2F(k,\theta)}{\sqrt{3}}\frac{\exp(ikr)}{r},
\end{equation} 
where $z$ is the direction of incidence, $\theta$ is the scattering angle, the
relative wavevector is defined as $2k/\sqrt{3}$, and $F(k,\theta)$ is the scattering
amplitude. 

For $r_+\rightarrow 0$ we have
$$\phi_0(r_+)\rightarrow \frac{1}{\sqrt{2\pi
a}}\left(\frac{1}{r_+}-\frac{1}{a}\right).$$
Then, comparing $\Psi$ (\ref{atomdimer.asympt}) with Eq.~(\ref{general.boundary}) we
obtain a relation between the functions $\chi$ and $f$:
\begin{equation} \label{chif}
\chi({\bf r})=\sqrt{a/8\pi}f({\bf r}).
\end{equation} 

As we commented in Section \ref{sec.general}, for $E<0$ we have to put $\Psi_0=0$ in
Eq.~(\ref{general.psi}). This leads to $D({\bf r})=0$, and Eq.~(\ref{general.main})
takes the form
\begin{equation}\label{atomdimer.main}
(\hat L_{E}-a^{-1}+\sqrt{-E})f({\bf r})=0.
\end{equation}
The links of our approach to the method of Skorniakov and Ter-Martirosian \cite{STM}
are seen from the fact that Eq.~(\ref{atomdimer.main}) leads to the same equation
for the Fourier transform of the function $f({\bf r})$ as  in Ref. \cite{STM}. 

We first demonstrate a  general approach for solving Eq.~(\ref{atomdimer.main}) and
finding the scattering amplitude. For $r\gg a$, the second term on the right hand
side of
Eq.~(\ref{general.L}) is exponentially small and can be omitted. Performing the
integration in the first term we reduce Eq.~(\ref{general.L}) to the form
\begin{eqnarray}  \label{L1}
&&\hat L_E f({\bf r})=\int {\rm d}^3q\left\{\sqrt{-E+q^2}-\sqrt{-E}\right\} \nonumber
\\
&&\times f({\bf q})\exp(i{\bf qr});\,\,\,\,\,\,\,\,r\gg a,
\end{eqnarray}
where $f({\bf q})$ is the Fourier transform of the function $f({\bf r})$. Acting
twice with the operator $(\hat L_E+\sqrt{-E})$ on the function $f({\bf r})$ and
using Eq.~(\ref{atomdimer.main}) we then obtain
\begin{equation} \label{L2}
(\hat L_E+\sqrt{-E})^2f({\bf r})=\int {\rm d}^3q(q^2-E)f({\bf q})e^{i{\bf
qr}}=\frac{f({\bf r})}{a^2}.
\end{equation}
The total energy is given by 
\begin{equation} \label{E}
E=-1/a^2+k^2,
\end{equation}
and Eq.~(\ref{L2}) immediately transforms into the equation for the free relative
atom-dimer motion;
\begin{equation}   \label{ffree}
(-\nabla_{{\bf r}}^2-k^2)f({\bf r})=0;\,\,\,\,\,\,\,\,\,\,r\gg a.
\end{equation}

The expansion of the function $f({\bf r})$ and the
amplitude $F(k,\theta)$ in Legendre polynomials reads (see Appendix): 
\begin{eqnarray} 
& &f({\bf r}) = \sum_{l=0}^\infty i^l(2l+1)P_l(\cos{\theta})f_l(r),
\label{fexpansion}  \\
& &F(k,\theta)=\sum_{l=0}^{\infty}(2l+1)P_l(\cos{\theta})F_l(k). \label{Aexpansion}
\end{eqnarray}
For the function $f_l(r)$, which describes the scattering with orbital angular
momentum $l$, the superposition of an incident and scattered waves satisfying
Eq.~(\ref{ffree}) can be written as
\begin{align} \label{fl}
&\sqrt{a/8\pi}f_l(r)=\sqrt{\pi/2kr}\{ J_{l+1/2}(kr) \nonumber \\
&+i(2k/\sqrt{3})F_l(k)H_{l+1/2}(kr)\};\,\,\,\,\,\,\,r\gg a, 
\end{align}
where $H_{l+1/2}$ is a Hankel function representing an outgoing wave, and $J_{l+1/2}$
is a Bessel function. For $kr\rightarrow\infty$ the Hankel function is
$H_{l+1/2}\simeq (-i)^{l+1}\sqrt{2/\pi kr}\exp(ikr)$. Then, multiplying both sides
of Eq.~(\ref{fl}) by $i^l(2l+1)P_l(\cos\theta)$, making a summation over $l$, and
taking into account Eqs.~(\ref{fexpansion}), (\ref{Aexpansion}) and (\ref{chif}), we
arrive at Eq.~(\ref{atomdimer.f}).  

Using Eq.~(\ref{fexpansion}) one reduces Eq.~(\ref{atomdimer.main}) to a set of
uncoupled integral equations for the functions $f_l(r)$:
\begin{equation}\label{atomdimer.mainl}
(\hat L_{E}^{l}-a^{-1}+\sqrt{-E})f_l(r)=0;\,\,\,\,\,\,l=0,1,2,...,
\end{equation}
where the integral operator $\hat{L}_{E}^{l}$ for a given $l$ is obtained by
integrating $\hat{L}_{E}$ over $dO_{{\bf r}}/4\pi$  with the weight
$P_l(\cos{\theta})$. Partial scattering amplitudes $F_l( k)$ are then found by
solving Eqs.~(\ref{atomdimer.mainl}) and fitting the obtained $f_l(r)$ with the
asymptotic expression (\ref{fl}). Then, equations (\ref{fexpansion}) and
(\ref{Aexpansion}) give the function 
$f({\bf r})$ and the total scattering amplitude $F(k,\theta)$.

In the ultracold limit, where the condition
\begin{equation} \label{ka}
ka\ll 1
\end{equation}
is satisfied, the scattering is dominated by the $s$-wave contribution and can be
analyzed setting $k=0$ and writing the total energy as $E=-1/a^2$. This  is  clearly
seen from Eq.~(\ref{fl}) with $l=0$, which for $k\rightarrow 0$ reads:
\begin{equation}\label{atomdimer.swave}
f_0(r)\approx \sqrt{8\pi/a}\,(1-2a_{ad}/\sqrt{3}r);\,\,\,\,\,\,\,\,\,r\gg a,
\end{equation}
where $a_{ad}=-F_0(0)$ is the atom-dimer scattering length. 

Thus, for finding $a_{ad}$ one should solve Eq.~(\ref{atomdimer.mainl}) with $l=0$,
assuming the limit $k\rightarrow 0$. This equation then reduces to
\begin{equation} \label{LE0}
\hat L_E^0f_0(r)=0.
\end{equation}
Using Eq.~(\ref{general.green}) and integrating over $dO_{{\bf r}}/4\pi$ in
Eq.~(\ref{general.L}) we represent Eq.~(\ref{LE0}) in the form:
\begin{eqnarray} \label{LE01}
&&\hat L_E^0f_0(r)=\frac{1}{\pi
ar}\int_0^{\infty}\Big[(f_0(r)-f_0(r'))\Big\{\frac{K_1(|r'-r|/a)}{|r'-r|} 
\nonumber \\
&&-\frac{K_1((r+r')/a)}{r+r'}\Big\}+2f_0(r')\Big\{\frac{K_1(\sqrt{r^2+r'^2-rr'}/a)}{\sqrt{r^2+r'^2-rr'}}
\nonumber \\
&&-\frac{K_1(\sqrt{r^2+r'^2+rr'}/a)}{\sqrt{r^2+r'^2+rr'}}\Big\}\Big]r'dr'=0,
\end{eqnarray}
where $K_1$ is an exponentially decaying Bessel function.
It is easily seen that the principal value of the integral in the first line of
Eq.~(\ref{LE01}) is finite. For $r'\rightarrow r$ the integrand behaves as
$1/(r-r')$.
 
We numerically solved Eq.~(\ref{LE01}) and found the function $f_0(r)$ at all
distances $r$. Fitting the obtained $f_0(r)$ at $r\gg a$ with the asymptotic
expression  (\ref{atomdimer.swave}) we arrived at the atom-dimer scattering length
$a_{ad}=1.2a$, which reproduces the result of Ref. \cite{STM}.
Our calculations  also show that the behavior of $f_0$ suggests a  soft-core
atom-dimer repulsion,
with a range of the order of $a$.

\section{Relaxation in atom-dimer collisions\label{sec.3bodyrel}}

Weakly bound dimers that we are considering are diatomic molecules in the 
highest ro-vibrational state. Hence, they can undergo relaxation into deep bound
states in
their collisions with each other or with unbound atoms. The released binding
energy of a deep state is $\sim\hbar^2/mR_e^2$. It is transformed into the kinetic
energy of particles in the outgoing collisional channel and they escape from the
trapped sample. Therefore, the process of collisional relaxation of weakly bound
dimers
determines  the lifetime of a gas of these molecules and, in particular,
possibilities to
Bose-condense such a gas. 

In our  previous work \cite{Petrov1} we have shown that this process is suppressed
due to Fermi 
statistics for the atoms in combination with a large size of the dimer. The binding
energy of
the dimers is $\varepsilon_0=\hbar^2/ma^2$ and their size is $\sim a\gg R_e$. The
size of deep bound states is of the order of $R_e$. Hence, the relaxation requires
the presence of at least three
fermions at distances $\sim R_e$ from each other. As two of them are necessarily
identical, due to the Pauli exclusion principle the relaxation probability acquires
a small factor proportional to a power of $(qR_e)$, where $q\sim 1/a$ is a
characteristic momentum of the atoms in the weakly bound molecular state. 

In this Section we discuss relaxation of weakly bound dimers into deep bound states
in ultracold atom-dimer collisions satisfying the condition (\ref{ka}) . Relying on
the inequality $a\gg R_e$ we develop a method that allows us to establish a
dependence of the relaxation rate on the scattering length $a$, without going into a
detailed analysis of the short-range behavior of
the systems of three atoms. It is assumed that the amplitude of the inelastic
process of relaxation is much smaller than the amplitude of elastic scattering. Then
the dependence of the relaxation rate on $a$ is related only to the $a$-dependence
of the initial-state 3-body wavefunction $\Psi$.

The relaxation occurs when all of the three atoms approach each other to distances of
the order of
$R_e$. At these interatomic distances, as well as at all distances $x,y\ll a$,  the
wavefunction
$\Psi$ in the ultracold limit is determined by the Schr\"odinger equation
(\ref{general.schrodinger}) with $E=0$. Therefore,  it depends on the scattering
length $a$ only through a
normalization coefficient: 
\begin{equation} \label{Psipsi}
\Psi=A(a)\psi;\,\,\,\,\,\,\,\,\,x,y\ll a,
\end{equation}
where the function $\psi$ is independent of $a$. 
The probability of relaxation and, hence, the relaxation rate constant $\alpha_{ad}$
are proportional to $|\Psi|^2$ at distances $x,y\sim R_e$. We thus  have
\begin{equation} \label{alphaA}
\alpha_{ad}\propto |A(a)|^2.
\end{equation}
The goal then is to find the coefficient $A(a)$, which determines the dependence of
the relaxation rate on $a$. 

In the region where $R_e\ll \{x,y\}\ll a$, the $a$-independent function $\psi$ can be
found in the  zero-range approximation. Then, matching the wavefunction $\Psi$ given
by Eq.~(\ref{Psipsi}), with the result 
of the zero range approximation at interparticle distances larger than $a$ gives the
coefficient $A(a)$.   

We start with analyzing the behavior of the 3-body wavefunction in the zero range
approximation at distances $x,y\ll a$. In this region of distances, $\Psi$ is
reconstructed through the function $f({\bf r})$ from Eq.~(\ref{general.psi}) with
$h({\bf r})=f({\bf r})$ and $\Psi_0=0$, setting $E\rightarrow 0$. 
Accordingly, one should use the Green function $G_0$ (\ref{G0}) in this equation,
which then reads:
\begin{align} \label{Psishort}
&\Psi({\bf x},{\bf y})=\int \Big[ G_0\Bigl(\!\sqrt{({\bf x}-{\bf r'}/2)^2+({\bf
y}+\sqrt{3}{\bf r'}/2)^2}\Bigr) \nonumber \\
& -\int G_0\Bigl(\!\sqrt{({\bf x}-{\bf r'}/2)^2+({\bf
y}-\sqrt{3}{\bf r'}/2)^2}\Bigr)\Big] \nonumber \\
&\times f({\bf r'}){\rm d}^3r';\,\,\,\,\,\,\,\,\,\,x,y\ll a.
\end{align} 
The main contribution to the integral in Eq.~(\ref{Psishort}) comes from distances
$r'\ll a$ and, hence, we have to find the function $f$ at these distances. In the
ultracold limit where the condition (\ref{ka}) is satisfied, the inelastic process
of relaxation is dominated by the contribution of the $s$-wave channel. As we found
in Section~\ref{sec.atommol}, the $s$-wave part $f_0$ of the function $f$ is
determined by
Eqs.~(\ref{LE0}) and (\ref{LE01}). For $r\ll a$, the distances $r'$ in
Eq.~(\ref{LE01}) are also much smaller than $a$. This is equivalent to setting the
limit $E\rightarrow 0$ ($a^{-1}\rightarrow 0$) for the integral operator $\hat
L_E^0$ in Eq.~(\ref{LE01}), and this operator then reduces to
\begin{align} \label{L00}
&\hat L_0^0f_0(r)=\frac{1}{\pi r}\int_0^{\infty}\Big[(f_0(r)-f_0(r')) \nonumber \\
&\times\Big\{\frac{1}{(r'-r)^2}-\frac{1}{(r+r')^2}\Big\}+2f_0(r') \nonumber \\
&\times\Big\{\frac{1}{r^2+r'^2-rr'}-\frac{1}{r^2+r'^2+rr'}\Big\}\Big]r'dr',
\end{align} 
where the integration of the term containing $(r-r')^2$ assumes a principal value of
the integral.

Thus, the function $f_0(r)$ at $r\ll a$ is a solution of the integral equation 
\begin{equation} \label{L000}
\hat L_0^0f_0(r)=0.
\end{equation}
The operator $\hat L_0^0$ has a property that 
$\hat{L}_0^0r^\nu=\lambda(\nu)r^{\nu-1}$ for $-5<{\rm Re}\,(\nu)<3$, and the integral
in
Eq.~(\ref{L00}) diverges outside this interval. The function $\lambda(\nu)$ is given
by
\begin{equation}\label{atomdimer.lambda0}
\lambda(\nu)=(\nu+1)\tan\frac{\pi\nu}{2}+\frac{4}{\sqrt{3}}\frac{\sin[\pi
(\nu+1)/6]}{\cos(\pi\nu/2)}.
\end{equation}
In the specified interval of $\nu$ the function $\lambda$ has two roots,
$\nu_+=1.166$ and $\nu_-=-3.166$.  Accordingly, the solution of
Eq.~(\ref{L000}) is a linear superposition: 
\begin{equation}\label{atomdimer.asym}
f_0(r)\approx C_+r^{\nu_+}+C_-r^{\nu_-},\,\,\,\,\,\,\,\,r\ll a. 
\end{equation}
The determination of the ratio $C_+/C_-$ involves short-range physics and is beyond
the scope of this paper. However, in the absence of a three-body resonance the
matching procedure implies that at distances $r\sim R_e$ both terms in
Eq.~(\ref{atomdimer.asym}) are of the same order of magnitude. So $C_+/C_-\propto
R_e^{-4.332}$, and at distances $r\gg R_e$ one has $f_0(r)\approx C_+r^{\nu_+}$. 
Substituting this result into Eq.~(\ref{Psishort}) we find that at distances
$x,y\ll a$ the three-body wavefunction takes the form
\begin{equation}\label{atomdimer.psiasym}
\Psi\approx
\Phi^0(\Omega)f(\rho)/\rho=C_+\Phi^0(\Omega)\rho^{\nu_+-1};\,\,\,\,\,\,\,\,x,y\ll a,
\end{equation}
where $\rho=\sqrt{x^2+y^2}$ is the hyperradius, and the set of hyperangles $\Omega$
denotes all the other coordinates. Although one can explicitly write down
the function $\Phi^0(\Omega)$, for us it is only important that this function is
$a$-independent.  

Comparing Eq.~(\ref{atomdimer.psiasym}) with Eq.~(\ref{Psipsi}) we see that one may
set
$\psi=\Phi^0(\Omega)\rho^{\nu_+-1}$ in the interval of distances where $R_e\ll
\{x,y\}\ll a$.
We then have $C_+=A(a)$, i.e. the function $f_0$ can be written as
\begin{equation} \label{atomdimer.asymfin} 
f_0(r)\approx A(a)r^{\nu_+},\,\,\,\,\,\,\,r\ll a.
\end{equation}
Numerical integration of Eq.~(\ref{LE01}) shows that $f_0$ smoothly
interpolates between the asymptotic expressions given by
Eq.~(\ref{atomdimer.asymfin}) 
for $r\ll a$, and by Eq.~(\ref{atomdimer.swave}) for $r\gg a$. This procedure
provides matching of the two asymptotes at $r\sim a$ and gives the coefficient
\begin{equation} \label{Aa}
A(a)\propto a^{-1/2-\nu_+}.
\end{equation}

In fact, the result of Eq.~(\ref{Aa}) can be obtained in a more elegant way, using
only 
Eqs.~(\ref{atomdimer.swave}) and (\ref{atomdimer.asymfin}). In the zero range
approximation the only distance scale is the 2-body scattering length $a$. Hence, we
may rescale the coordinate and represent the function $f_0$ in the form
\begin{equation} \label{frescaled}
f_0(r)=B(a)\tilde f_0(r/a),
\end{equation}
where $\tilde f_0$ depends on $a$ only through the rescaled coordinate $r/a$. The
coefficient $B(a)$
is independent of the coordinate and can be obtained by comparing
Eq.~(\ref{frescaled}) with 
asymptotic expressions (\ref{atomdimer.swave}) and (\ref{atomdimer.asymfin}). Using 
Eq.~(\ref{atomdimer.swave}) we see that $B\propto a^{-1/2}$, whereas the comparison
of Eq.~(\ref{frescaled}) with Eq.~(\ref{atomdimer.asymfin}) gives $B\propto
A(a)a^{\nu_+}$. This immediately leads to Eq.~(\ref{Aa}) for the coefficient $A(a)$. 
  
As the dependence of the relaxation rate constant $\alpha_{ad}$ on the 2-body
scattering length is governed by Eq.~(\ref{alphaA}), using Eq.~(\ref{Aa}) we obtain
$\alpha_{ad}\propto a^{-s}$, where $s=1+2\nu_+\simeq 3.33$. The absolute value of
the relaxation rate is determined by the contribution of interparticle distances
$\sim\!R_e$, where the zero-range approximation is not valid. This approximation only
gives a
correct dependence of the relaxation rate on $a$.

Assuming that the short-range physics is characterized by the length scale $R_e$ and
by the energy scale $\hbar^2/mR_e^2$, we can restore the dimensions and write the
following expression for the rate constant of relaxation of weakly bound dimers into
deep bound states in atom-dimer ultracold collisions: 
\begin{equation} \label{alphaatom}
\alpha_{ad}= C (\hbar R_e/m)(R_e/a)^s;\,\,\,\,\,\,\,\,s=3.33.
\end{equation}
One clearly sees that the relaxation rate rapidly decreases with increasing the
2-body scattering length.
However, the coefficient $C$ depends on a particular system.

The relaxation due to atom-dimer scattering with non-zero orbital angular momenta is
very small.
Since the atom-dimer effective interaction has a characteristic range $\sim a$ (see
Sec.~\ref{sec.atommol}), the $p$-wave part of the 3-body wavefunction $\Psi$ at
short interparticle distances is proportional to $ka$. Hence, the $p$-wave
contribution to the relaxation rate is 
$\propto (ka)^2$ and can be omitted for ultracold collisions satisfying the condition
(\ref{ka}). 

\section{Elastic dimer-dimer scattering}

As we already mentioned in Introduction, elastic interaction between weakly bound
bosonic molecules of $\uparrow$ and $\downarrow$ fermionic  atoms is important for
understanding the physics of their Bose-Einstein condensation and for studying the
BCS-BEC cross-over in two-component atomic Fermi gases. The dimer-dimer elastic
scattering is a 4-body problem described by the Schr\"odinger equation
\begin{eqnarray}\label{4bodySchrU}
&&\Big\{-\nabla_{{\bf r}_1}^2-\nabla_{{\bf r}_2}^2-\nabla_{\bf
{R}}^2+U(r_1)+U(r_2) \nonumber \\
&&+\sum_{\pm}U[({\bf r}_1+{\bf r}_2\pm\sqrt{2}{\bf R})/2]-E\Big\}\Psi=0.
\end{eqnarray}
Here we again use units in which $\hbar=m=1$. The distance between two given
$\uparrow$ and 
$\downarrow$ fermions is ${\bf r}_1$, and ${\bf r}_2$ is the distance between the
other two.
The distance between the centers of mass of these pairs is ${\bf {R}}/\sqrt{2}$, and
$({\bf r}_1+{\bf r}_2
\pm\sqrt{2}{\bf R})/2$ are the separations between $\uparrow$ and $\downarrow$
fermions in the other two possible $\uparrow\downarrow$ pairs. The total energy is
$E=-2\varepsilon_0+\varepsilon$, with
$\varepsilon$ being the collision energy, and $\varepsilon_0=-1/a^2$ the binding
energy of a dimer.

The wavefunction $\Psi$ is symmetric with respect to the
permutation of bosonic $\uparrow\downarrow$ pairs and antisymmetric with respect to
permutations of identical fermions:
\begin{align}\label{symmetry}
&\!\!\Psi({\bf r}_{1},{\bf r}_{2},{\bf {R}})=
\Psi({\bf r}_{2},{\bf r}_{1},-{\bf {R}}) \nonumber  \\
&=-\Psi\left(\frac{{\bf r}_1+{\bf r}_2 \pm\sqrt{2}{\bf R}}{2},\frac{{\bf r}_1+{\bf
r}_2 \mp\sqrt{2}{\bf R}}{2}, \pm\frac{{\bf r}_1-{\bf r}_2}{\sqrt{2}}\right).
\end{align}

The weak binding of atoms in the dimer assumes that the 2-body (positive) scattering
length is 
$a\gg R_e$, and we employ the zero range approximation in our analysis of the
molecule-molecule scattering. This is done relying on the formulation of this
approximation given in Sec.~\ref{sec.general} for the three-body problem. Thus, the
4-body system is described by the free-particle Schr\"odinger equation
\begin{equation}\label{4bodySchr}
-\left[\nabla_{{\bf r}_1}^2+\nabla_{{\bf r}_2}^2+\nabla_{\bf
{R}}^2+E\right]\Psi=0,
\end{equation}
and the 4-body wavefunction $\Psi$ should satisfy the Bethe-Peierls boundary
condition for a vanishing distance in any pair of $\uparrow$ and $\downarrow$
fermions, i.e. for ${\bf r}_1\rightarrow 0$, ${\bf r}_2\rightarrow 0$, and ${\bf
r}_1+{\bf r}_2 \pm\sqrt{2}{\bf R}\rightarrow 0$. Due to the symmetry it is necessary
to require a proper behavior of $\Psi$ only at one of these boundaries. For ${\bf
r}_1\rightarrow 0$ the boundary condition reads:
\begin{equation}\label{boundary}
\Psi({\bf r}_1,{\bf r}_2,{\bf {R}})\rightarrow f({\bf r}_2,{\bf
R})(1/4\pi r_1\,-1/4\pi a).
\end{equation}
The function $f({\bf r}_2,{\bf R})$ is analogous to that defined in
Sec.~\ref{sec.general} and it contains the information about the second pair of
particles when the first two are sitting on top of
each other.

In the ultracold limit, where the condition (\ref{ka}) is satisfied, the scattering
is dominated by the
contribution of the $s$-wave channel. The inequality (\ref{ka}) is equivalent to
$\varepsilon\ll \varepsilon_0$ and, hence, the $s$-wave scattering  can be analyzed
from the solution of
Eq.~(\ref{4bodySchr}) with $E=-2\varepsilon_0<0$. For large $R$ the corresponding
wavefunction
is given by
\begin{equation}   \label{asymptote}
\Psi\approx\phi_0(r_1)\phi_0(r_2)(1-\sqrt{2}a_{dd}/R);\,\,\,\,\,R\gg a,
\end{equation}
where $a_{dd}$ is the dimer-dimer scattering length, and the wavefunction of the
weakly bound dimer is given by Eq.~(\ref{twobody.boundstate}). 
Combining Eqs.~(\ref{boundary}) and (\ref{asymptote}) we obtain the asymptotic
expression for 
$f$ at large distances $R$:
\begin{equation}\label{dimerdimer.swave}
f({\bf r}_2,{\bf R})\approx (2/r_2a)\exp{(-r_2/a)}(1-\sqrt{2}a_{dd}/R);\,\,\,\,R\gg
a.
\end{equation}

In the case of the $s$-wave scattering the function $f$ depends only on
three variables: the absolute values of ${\bf r}_2$ and ${\bf {R}}$, and the angle
between them. We now derive and solve the equation for $f$. The value of $a_{dd}$ is
then deduced from the behavior of $f$ at large $R$ governed by
Eq.~(\ref{dimerdimer.swave}).

We first establish a general form of the wavefunction $\Psi$ satisfying
Eq.~(\ref{4bodySchr}), with the boundary condition (\ref{boundary}) and symmetry
relations (\ref{symmetry}). In our case the total energy $E=-2/a^2<0$, and the Green
function of Eq.~(\ref{4bodySchr}) reads
\begin{equation} \label{Green4body}
G(X)=(2\pi)^{-9/2}(Xa/\sqrt{2})^{-7/2} K_{7/2}(\sqrt{2}\,X/a),
\end{equation}
where $X=|S-S'|$, and $S=\{{\bf r}_1,{\bf r}_2,{\bf R}\}$ is a 9-component vector.
Accordingly, one has
$|S-S'|=\sqrt{({\bf r}_1-{\bf r'}_1)^2+({\bf r}_2-{\bf r'}_2)^2+({\bf R}-{\bf
R'})^2}$. In analogy with the 3-body case, the 4-body wavefunction $\Psi$ can be
expressed through $G(|S-S'|)$ with coordinates $S'$ corresponding to a vanishing
distance between $\uparrow$ and $\downarrow$ fermions, i.e. for 
${\bf r'}_1\rightarrow 0$, ${\bf r'}_2\rightarrow 0$, and $({\bf r'}_1+{\bf
r'}_2\pm\sqrt{2}{\bf R'})/2\rightarrow 0$. Thus, for the wavefunction satisfying the
symmetry relations (\ref{symmetry}) we have
\begin{eqnarray}\label{Psi4body}
&&\Psi(S)=\Psi_0+\int d^3r'd^3R'\Big[ G(|S-S_1|)+G(|S-S_2|) \nonumber \\
&&-G(|S-S_+|)-G(|S-S_-|)\Big]h({\bf r'},{\bf R'}),
\end{eqnarray}
where $S_1=\{0,{\bf r}',{\bf {R}}'\},\;S_2=\{{\bf r}',0,-{\bf {R}}'\}$, and
$S_\pm=\{{\bf r}'/2\pm{\bf {R}}'/\sqrt{2},{\bf r}'/2\mp{\bf {R}}'/\sqrt{2},\mp{\bf
r}'\sqrt{2}\}$.
The function $\Psi_0$ is a properly symmetrized finite solution of
Eq.~(\ref{4bodySchr}),
regular at vanishing distances between $\uparrow$ and $\downarrow$ fermions. For
$E<0$, non-trivial solutions of this type do not exist and we have to put
$\Psi_0=0$. The function $h({\bf r}_2,{\bf R})$ has to be determined by comparing
$\Psi$ (\ref{Psi4body}) at ${\bf r}_1\rightarrow 0$, with the boundary condition
(\ref{boundary}).   

This procedure is similar to that developed in Section~\ref{sec.general} for the
3-body case. Considering the limit ${\bf r}_1\rightarrow 0$ we extract the leading
terms on the right hand side of Eq.~(\ref{Psi4body}). These are the terms that behave
as $1/r_1$
or remain finite in this limit. The last three terms in the square brackets in
Eq.~(\ref{Psi4body}) provide a finite contribution
\begin{align} \label{reg1}
&\int d^3r'd^3R'\,h({\bf r}',{\bf R}')\Big[G(|{\bar S}_2-S_2|) \nonumber \\
&-G(|{\bar S}_2-S_+|)-G(|{\bar S}_2-S_-|)\Big],
\end{align}    
where ${\bar S}_2=\{0,{\bf r}_2,{\bf R}\}$. For finding the  contribution of the
first term in the square brackets we subtract from this term and add to it an
auxiliary quantity
\begin{eqnarray} \label{aux0}
&&h({\bf r}_2,{\bf R})\int G(|S-S_1|)d^3r'd^3R'= \nonumber \\
&&=\frac{h({\bf r}_2,{\bf R})}{4\pi r_1}\exp{(-\sqrt{2}r_1/a)}.
\end{eqnarray}
The result of the subtraction yields a finite contribution which for $r_1\rightarrow
0$ can be written as
\begin{align} \label{reg2}
&\int d^3r'd^3R'[h({\bf r}',{\bf R}')-h({\bf r}_2,{\bf R})]G(|S-S_1|)=  \nonumber \\
&=P\int d^3r'd^3R'[h({\bf r}',{\bf R}')-h({\bf r}_2,{\bf R})]G(|{\bar S}_2-S_1|),
\end{align}
with the symbol $P$ standing for the principal value of the integral over $dr'$ (or
$dR'$). 
Equation (\ref{reg2}) is derived in Appendix and it is proven that the integral in
the second line of this equation is  convergent. 

In the limit $r_1\rightarrow 0$, the right hand side of Eq.~(\ref{aux0}) is equal to
\begin{equation} \label{aux}
h({\bf r}_2,{\bf R})(1/4\pi r_1-\sqrt{2}/4\pi a).
\end{equation}
We thus  find that for ${\bf r}_1\rightarrow 0$ the wavefunction $\Psi$ of
Eq.~(\ref{Psi4body}) takes the form
\begin{equation} \label{Psilim}
\Psi({\bf r}_1,{\bf r}_2,{\bf R})=\frac{h({\bf r}_2,{\bf R})}{4\pi r_1}+{\cal
R};\,\,\,\,\,\,{\bf r}_1\rightarrow 0,
\end{equation}
where ${\cal R}$ is the sum of regular $r_1$-independent terms given by
Eqs.~(\ref{reg1}) and (\ref{reg2}), and by the second term on the right hand side of
Eq.~(\ref{aux}). Equation (\ref{Psilim}) should coincide
with Eq.~(\ref{boundary}), and comparing the singular terms of these equations we
find $h({\bf r}_2,{\bf R})=f({\bf r}_2,{\bf R})$. As  the quantity ${\cal R}$ should
coincide with the regular term of Eq.~(\ref{boundary}), equal to $-f({\bf r}_2,{\bf
R})/4\pi a$, we obtain the following equation for the function $f$:
\begin{eqnarray}\label{main}
&&\int d^3r'd^3R'\Big\{G(|{\bar S}-S_1|)[f({\bf r}',{\bf {R}}')-f({\bf r},{\bf {R}})]
\nonumber\\
&&+\bigl[G(|{\bar S}-S_2|)-\sum_\pm G(|{\bar S}-S_\pm|)\bigr]f({\bf r}',{\bf
{R}}')\Big\} \nonumber \\
 &&=(\sqrt{2}-1)f({\bf r},{\bf {R}})/4\pi a.
\end{eqnarray}
Here ${\bar S}=\{0,{\bf r},{\bf {R}}\}$, and we omitted the symbol of principal value
for the integral in the first line of Eq.~(\ref{main}). 

As we already mentioned above, for the $s$-wave scattering the function $f({\bf
r},{\bf R})$ depends only on the absolute values of ${\bf r}$ and ${\bf R}$ and on
the angle between them. Thus, Eq.~(\ref{main}) is an integral equation for the
function of three variables. We have solved this equation numerically for all
distances $R$ and $r$, and all angles between the vectors ${\bf R}$ and ${\bf r}$.
Fitting the asymptotic expression (\ref{dimerdimer.swave}) at $R\gg a$ with the
function $f({\bf r},{\bf R})$ obtained numerically from Eq.~(\ref{main}), we find
with 2\% accuracy that the dimer-dimer scattering length
is:
\begin{equation}   \label{add}
a_{dd}=0.6 a>0.
\end{equation}
Our calculations show the absence of 4-body weakly bound states, and the behavior of
$f$ at small $R$ suggests a soft-core repulsion between dimers, with a range $\sim
a$. 

The result of Eq.~(\ref{add}) is exact, and its derivation was outlined in our
previous work \cite{Petrov1}. Equation (\ref{add}) indicates the stability of
molecular BEC with respect to collapse. Compared to earlier studies which assumed
$a_{dd}=2a$ \cite{Rand}, Eq.~(\ref{add}) gives almost twice as small a sound
velocity of the molecular condensate and a rate of elastic collisions smaller by an
order of magnitude. We should mention here that the result of earlier studies
\cite{Rand} was reconsidered in Ref. \cite{strinati} by using a diagrammatic
approach which leads to $a_{dd}=0.75a$. However, this approach misses a number of
diagrams which give a contribution of the same order of magnitude as those taken
into account.

\section{Relaxation in dimer-dimer collisions}

In this Section we generalize the results obtained in Sec.~\ref{sec.3bodyrel} to the
relaxation of weakly bound dimers into deep bound states in dimer-dimer collisions.
We again consider the ultracold limit described by the condition (\ref{ka}), where
the relaxation process is dominated by the contribution of the $s$-wave dimer-dimer
scattering. The key point of our discussion is that the dimer-dimer relaxation
collisions are to a large extent similar to the atom-dimer ones. 

Indeed, the relaxation process requires only three atoms to approach each other to
short distances of the order of $R_e$. The fourth particle can be at a large
distance from these three and, in this respect, does not participate in the
relaxation process. This distance is of the order of the size of a dimer, which is
$\sim a\gg R_e$. As well as in the case of atom-dimer collisions, the dependence of
the relaxation rate on the 2-body scattering length $a$ is determined by the
$a$-dependence of the initial state wavefunction $\Psi$. We thus see that the
configuration space contributing to the relaxation probability can be viewed as a
system of three atoms at short distances $\sim R_e$ from each other and a  fourth
atom separated from this system by a large distance $\sim a$. In this case the
4-body wavefunction decomposes into a product: 
\begin{equation}\label{decomp}
\Psi=\eta({\bf z})\Psi^{(3)}(\rho,\Omega),
\end{equation}
where  $\Psi^{(3)}$ is the wavefunction of the 3-fermion system, $\rho$ and $\Omega$
are the hyperradius and the set of hyperangles for these fermions,  ${\bf z}$ is the
distance between their center of mass and the fourth atom, and the function
$\eta({\bf z})$ describes the motion of this atom. Note that Eq.~(\ref{decomp})
remains valid for any hyperradius $\rho\ll |{\bf z}|\sim a$. 

The transition to a deep bound 2-body state occurs in the system of three atoms and
does not change the wavefunction of the fourth atom, $\eta({\bf z})$. Therefore,
averaging the transition probability over the motion of the fourth particle, the
rate constant of relaxation in dimer-dimer collisions can be written as  
\begin{equation} \label{alpha34}
\alpha_{dd}=\alpha^{(3)}\int |\eta({\bf z})|^2d^3z=\alpha^{(3)},
\end{equation}
where $\alpha^{(3)}$ is the rate constant of relaxation for the 3-atom system.

We thus obtain that the problem is reduced to the relaxation in atom-dimer
collisions. The difference from the case discussed in Section \ref{sec.3bodyrel} is
that now the relative momentum of the collision is $\sim 1/a$. This is seen by
considering $\Psi^{(3)}$ at large distances between the dimer and the fermionic atom,
hereafter referred to as the 3-rd fermion. This fermion is in the bound molecular
state
with the fourth atom. As the size of this state is $\sim a$, the expansion of the
wavefunction of the 3-rd fermion in plane waves shows that its momentum $q$ is of
the order of $1/a$. 
 
Keeping in mind the discussion in Section~\ref{sec.3bodyrel} we see that the result
for the relaxation rate following from Eq.~(\ref{alphaatom}) remains valid for the
dimer-dimer collisions. The fact that the relative momentum is $q\sim 1/a$ can only
change the numerical coefficient, not the dependence of the relaxation rate constant
on the 2-body scattering length $a$. However, the result of Eq.~(\ref{alphaatom})
accounts only for the $s$-wave scattering of the 3-rd fermionic atom on the dimer,
which provides  the leading relaxation channel for ultracold atom-dimer collisions.
In the case of dimer-dimer collisions, there is a relaxation channel that is more
important in the limit of large $a$ (see below). For the $s$-wave dimer-dimer
scattering, both the 4-th and the 3-rd fermions (bound to each other in the
molecular state) can undergo the $p$-wave scattering on the other dimer in such a
way that their total orbital angular momentum is equal to zero. We thus should
consider the relaxation for the $p$-wave collisions between the 3-rd fermionic atom
and the dimer. These collisions are not suppressed as their relative momentum is
$q\sim 1/a$.    

As we discussed in Section~\ref{sec.3bodyrel}, for the hyperradius $\rho\ll a$ the
wavefunction of the 3-fermion system, $\Psi^{(3)}$, depends on the scattering length
$a$ only through a normalization coefficient and can be written in the form
(\ref{Psipsi}). We thus have $\Psi^{(3)}=A(a)\psi$, with $a$-independent function
$\psi$, and the relaxation rate constant depends on $a$ as $\alpha_{dd}\propto
|A(a)|^2$. This requires us to find the coefficient $A(a)$ for the case of the
$p$-wave scattering of the 3-rd fermionic atom on the molecule. We  will do this  in
the zero range approximation, along the lines of our discussion of the $s$-wave
atom-dimer scattering in Section~\ref{sec.3bodyrel}.  

We first consider the region of interparticle distances where $R_e\ll\rho\ll a$. Then
the function $f({\bf r})$ for the 3-body problem is determined by
Eq.~(\ref{atomdimer.main}) in which we have to set the limit $E\rightarrow 0$ and
$a^{-1}\rightarrow 0$. Therefore, the integral equation (\ref{atomdimer.mainl}) for
the $p$-component of the function $f$ takes the form:
\begin{equation} \label{L01}
\hat L^1_0f_1(r)=0,
\end{equation}
where the operator $\hat L^1_0$ is obtained substituting the expansion
(\ref{fexpansion}) into Eq.~(\ref{general.L}) with the Green function $G_0$
(\ref{G0}), multiplying by $\cos{\theta}$, and integrating over the angles. This
yields
\begin{align} \label{L10}
&\hat
L_0^1f_1(r)=\frac{1}{\pi}\int_0^{\infty}dr'\Big(\frac{r'}{r}\Big[(f_1(r)-f_1(r'))\Big\{\frac{1}{(r'-r)^2}\nonumber
\\
&-\frac{1}{(r+r')^2}\Big\}+2f_1(r')\Big\{\frac{1}{r^2+r'^2-rr'}+\frac{1}{r^2+r'^2+rr'}\nonumber
\\
&-\frac{1}{(r+r')^2}\Big\}\Big]+\frac{2f_1(r')}{r^2}\ln{\frac{\sqrt{(r+r')}(r^2+r'^2-rr')}
{\sqrt{|r-r'|}(r^2+r'^2+rr')}}\Big),
\end{align} 
where one should take a principal value for the integral of the term containing
$(r-r')^2$.

Like the operator $L^0_0$, the operator $L^1_0$ has a property
$\hat{L}_0^1r^\nu=\lambda(\nu)r^{\nu-1}$ for $-4<{\rm Re}\,(\nu)<2$. The function
$\lambda(\nu)$ is now given by
\begin{align} \label{lambda1eq}
&\lambda(\nu)=\frac{4}{\sqrt{3}}\frac{\nu\cos\left[\pi(\nu+1)/6\right]-2\sin(\pi\nu/6)}{(\nu+1)\sin(\pi\nu/2)}\nonumber
\\
&+\frac{\nu(\nu+2)}{\nu+1}\cot\frac{\pi\nu}{2}
\end{align}
and has two roots: $\nu_+=0.773$ and $\nu_-=-2.773$. Then, with the same arguments as
in between equations (\ref{atomdimer.lambda0}) and (\ref{atomdimer.asymfin}) in
Section~\ref{sec.3bodyrel}, we obtain short-distance expressions for the function
$f_1$ and for the part of the 3-body wavefunction $\Psi^{(3)}$ corresponding to the
$p$-wave scattering of the 3-rd fermion on the molecule:
\begin{eqnarray} 
&&f_1(r)\approx A(a)a^{\nu_+}(r/a)^{\nu_+},\,\,\,\,\,\,\,\,r\ll a;  \label{f1nu} \\
&&\Psi^{(3)}=A(a)\Phi^1(\Omega)\rho^{\nu_+-1},\,\,\,\,\,\rho\ll a, \label{Psi3nu}
\end{eqnarray}
where the function $\Phi^1$ is  independent of $a$.

For relative momenta $k\sim 1/a$, the 2-body scattering length remains the only
distance scale of the zero range approximation. In particular, the $p$-wave
scattering amplitude in Eq.~(\ref{fl}) will be $F_1\sim a$. We therefore can write
the  function $f_1$ in the form (\ref{frescaled}): $f_1(r)=B(a)\tilde f_1(r/a)$,
where $\tilde f_1$ depends on $a$ only through the rescaled coordinate $r/a$. In
order to be consistent with Eq.~(\ref{fl}) for $k\sim 1/a$, we have to put
$B(a)\propto a^{-1/2}$. Then the comparison of the resulting $f_1$ with
Eq.~(\ref{f1nu}) gives the coefficient $A(a)\propto a^{-1/2-\nu_+}$ and the
relaxation rate constant is $\alpha_{dd}\propto |A|(a)|^2\propto a^{-s}$, with
$s=1+2\nu_+\simeq 2.55$. Restoring the dimensions we have
\begin{equation} \label{alphadd}
\alpha_{dd}=C(\hbar R_e/m)(R_e/a)^s;\,\,\,\,\,\,\,\,\,s=2.55.
\end{equation}

One can think of the relaxation mechanism, where the scattering of the 3-rd fermion
on the dimer occurs with higher orbital angular momenta $l$. In this case the 4-th
atom scatters on the dimer with the same $l$, and the total angular momentum of the
dimer-dimer scattering should be equal to zero. Our analysis shows that these
scattering mechanisms lead to a power law dependence $\alpha_{dd}\propto a^{-s}$,
with larger values of $s$ than in Eq.~(\ref{alphadd}). Hence, for large $a$ they can
be neglected.  

Equation (\ref{alphadd}) shows a slower decrease of the relaxation rate with
increasing $a$ than in the case of atom-dimer collisions.  Obviously, in the limit
$R_e/a\rightarrow 0$ the dimer-dimer
relaxation should dominate over the atom-dimer one. The competition between these two
relaxation processes can be present if the  ratio $R_e/a$ is  not too small and the
densities of dimers and atoms in the gas are comparable with each other.

\section{Wide and narrow Feshbach resonances}  \label{resonance}

In experiments with alkali atom gases large values of the 2-body scattering length
are  achieved by using Feshbach resonances. In the vicinity of the resonance, the
2-body problem is characterized by a strong coupling between the continuum states of
colliding atoms and a bound molecular state of another hyperfine domain of these
atoms. The resulting scattering length depends on the detuning from the resonance,
i.e. on the energy difference $\Delta$ between the border of the continuum of
colliding atoms and the bound molecular state. The splitting between the two
hyperfine domains and, hence, the detuning $\Delta$ depend on the magnetic field,
which makes the scattering length tunable by varying the field. 

One thus has a two-channel problem which can be described in terms of Breit-Wigner
scattering \cite{Breit,LL3}, the open channel being the states of colliding atoms
and the closed channel the bound molecular state of the other hyperfine domain.
Various aspects of this type of problems have been discussed by Feshbach
\cite{Feshbach} and Fano \cite{Fano}. In cold atom physics the idea of Feshbach
resonances was introduced in Ref. \cite{Verhaar}, and optically induced resonances
have been discussed in Refs. \cite{Gora,Bohn}. 

We now  analyze to which extent our results for 3-atom and 4-atom systems of fermions
with a positive 2-body scattering length $a\gg R_e$ describe the situation of
Feshbach resonances. For low collision energies $\varepsilon$, omitting the (small)
background scattering length, the scattering amplitude is given by \cite{LL3} :
\begin{equation} \label{FE}
F(\varepsilon)=-\frac{\hbar\gamma/\sqrt{m}}{\varepsilon+\Delta+i\gamma\sqrt{\varepsilon}},
\end{equation}
where the quantity $\hbar\gamma/\sqrt{m}\equiv W$ characterizes the coupling between
the two hyperfine domains. In Eq.~(\ref{FE}) the detuning $\Delta$ is positive if
the bound molecular state is below the continuum of colliding atoms. Then for
$\Delta>0$ we have a positive scattering length near the resonance,
$a=-F(0)=W/\Delta$. Introducing a characteristic length 
\begin{equation} \label{RW}
R^*=\hbar^2/mW 
\end{equation}
and expressing the scattering amplitude through the relative momentum of particles
$k=\sqrt{m\varepsilon}/\hbar$, equation (\ref{FE}) takes the form:
\begin{equation} \label{Fk}
F(k)=-\frac{1}{a^{-1}+R^*k^2+ik},
\end{equation}
The validity of Eq.~(\ref{Fk}) does not require the inequality $kR^*\ll 1$. At the
same time,  Eq.~(\ref{Fk}) formally coincides with the amplitude of scattering of
slow particles by a potential with the same scattering length $a$ and an effective
range $R=-2R^*$, valid under the condition $kR\ll 1$. 

The length $R^*$ is an intrinsic parameter of the Feshbach resonance problem. It
characterizes the width of the resonance. From Eqs.~(\ref{FE}) and(\ref{RW}) we see
that large $W$ and, hence, small $R^*$ correspond to a wide resonance, whereas small
$W$ and large $R^*$ lead to a narrow resonance. The issue of wide and narrow
resonances is now actively being discussed in literature
\cite{Bruun,Bruun2,Petrovbosons,Palo,Eric,Ho3}. We would like to point out here that
the use of terms ``wide'' and ``narrow'' depends on the problem under consideration.
For example, in the unitarity limit where $a\rightarrow\pm\infty$, Eq.~(\ref{Fk})
shows  that the length $R^*$ drops out of the problem under the condition $kR^*\ll
1$. In a quantum degenerate Fermi gas the characteristic momentum of particles is
the Fermi momentum $k_F=(3\pi^2n)^{1/3}$, where $n$ is  the gas density. Thus, the
inequality $k_FR^*\ll 1$, referred to as the condition of a wide resonance, ensures
universality of the problem \cite{Bruun,Bruun2,Palo,Eric,Ho3}. The only length and
energy scales in the gas will be the mean interparticle separation $n^{-1/3}$ and
the Fermi energy  $E_F=(3\pi^2n)^{2/3}\hbar^2/2m$, and the system acquires 
universal thermodynamics \cite{Ho1}.

In our case the situation is different. We are considering weakly bound diatomic
molecules in the open channel with a binding energy (\ref{twobody.benergy}), and our
discussion of atom-dimer and dimer-dimer collisions assumes that there is a weakly
interacting gas of these dimers and atoms. The most important limitation is related
to the binding energy and the wavefunction of the dimers. The energy of the weakly
bound molecular state is determined by the pole  of the scattering amplitude
(\ref{Fk}). One then finds that this state exists only for $a>0$ and the universal
expression (\ref{twobody.benergy}) used in our calculations,
$\varepsilon_0=\hbar^2/ma^2$, requires the inequality \cite{Petrovbosons}
\begin{equation} \label{Ra}
R^*\ll a.
\end{equation}
Under this condition the wavefunction of the weakly bound molecular state has only a
small admixture of the closed channel \cite{Petrovbosons}. 

Our calculations for the atom-dimer and dimer-dimer collisions were done in the
ultracold limit 
where $ka\ll 1$. In a thermal gas of atoms and dimers the characteristic momentum is
the thermal momentum $k_T=(2mT/\hbar^2)^{1/2}$, whereas for a degenerate, in
particular Bose-condensed gas of the dimers, the characteristic momentum is the
inverse healing length, $(na)^{1/2}$. In both cases the inequality $ka\ll 1$ assumes
the weakly interacting regime, where $na^3\ll 1$. One  can also see that in the
limit $ka\ll 1$ the inequality Eq.~(\ref{Ra}) makes the scattering
amplitude $F(k)$ (\ref{Fk}) momentum independent and equal to $-a$.  This justifies
the  use of our single-channel zero range approximation for calculating atom-dimer
and dimer-dimer interactions and collisional properties. We thus obtain that for our
problem the condition of a wide Feshbach resonance is given by Eq.~(\ref{Ra}). Under
this condition the problem is universal in the sense that the size of weakly bound
dimers, and atom-dimer and dimer-dimer scattering properties are characterized by a
single parameter, the 2-body scattering length $a$.

Note that the interaction between the two channels of the  Feshbach problem is
efficient at interparticle distances which are of the order of $R_e$. Therefore, the
Feshbach character of scattering does not influence the condition $a\gg R_e$ that
allows us to use the zero range approximation for the entire region of interparticle
distances.

\section{Concluding remarks} \label{conclusions}

In most experiments with weakly bound diatomic molecules produced by using Feshbach
resonances in a degenerate two-component atomic Fermi gas, the wide resonance
condition (\ref{Ra}) was satisfied. This was the case with $^{40}$K$_2$ molecules at
JILA \cite{jila1,jila2,jila3}, and with $^6$Li$_2$ at Innsbruck
\cite{rudy1,rudy2,rudy3,rudy4}, MIT \cite{mit1,mit2}, Duke \cite{duke1,duke2}, ENS
\cite{Bourdel,ens1,ens2}, and Rice \cite{randy2}. In these experimental studies the
length $R^*$ is of the  order of or smaller than 20\AA, and for the achieved values
of the scattering length $a$ from 500 to 2000\AA$\,$ the ratio $R^*/a$ is smaller
than $0.1$. The
only exception is the experiment at Rice with $^6$Li near a narrow Feshbach
resonance at 543 G \cite{randy1}. For this resonance the length $R^*$ is large and
at obtained values of $a$ the condition (\ref{Ra}) is not fulfilled. Therefore, the
Rice experiment \cite{randy1} cannot be described by our theory. 

Experimental studies of  dimers produced in a Fermi gas by using wide Feshbach
resonances at JILA, Innsbruck, MIT, and ENS are well described within our
theoretical approach. It should be mentioned that for $a>0$ and equal concentrations
of the two atomic components of the gas, at temperatures well below $E_F$
practically all atoms should be converted into dimers if the gas density satisfies
the inequality $na^3\ll 1$ \cite{Servaas}. In ongoing experiments the imbalance
between the atomic components is fairly small, and at sufficiently low temperatures
there can only be a small fraction of unpaired fermionic atoms. 

The results at JILA \cite{jila1,jila2,jila3}, Innsbruck \cite{rudy1,rudy2,rudy3}, MIT
\cite{mit1,mit2}, ENS \cite{ens1,ens2}, and Rice \cite{randy2} show a remarkable
collisional stability of weakly bound diatomic molecules $^{40}$K$_2$ and
$^6$Li$_2$. At molecular densities $n\sim 10^{13}$ cm$^{-3}$ the lifetime of the gas
ranges from tens of milliseconds to tens of seconds, depending on the value of the
scattering length $a$. A strong decrease of the relaxation rate with increasing $a$,
following from Eq.~(\ref{alphadd}), is  consistent with experimental data. The
potassium experiment at JILA \cite{jila1} and lithium experiment at ENS \cite{ens2}
give the relaxation rate constant $\alpha_{dd}\propto a^{-s}$, where $s\approx 2.3$ 
with 15\% of accuracy for K$_2$, and $s\approx 2.0$ with 40\% of accuracy for Li$_2$. 
The absolute value of the rate constant
for a thermal gas of Li$_2$ is $\alpha_{dd}\approx 2\times 10^{13}$ cm$^3$/s for the
scattering length $a\approx 800$\AA $\,$ \cite{ens1}. For K$_2$ it is by an order of
magnitude higher at the same value of $a$ \cite{jila1}, which can be a consequence
of a larger value of the characteristic radius of interaction $R_e$.

At realistic temperatures the relaxation rate constant $\alpha_{dd}$ is much smaller
than the rate constant of elastic collisions $8\pi a_{dd}^2v_T$, where $v_T$ is the
thermal velocity. For example, for the Li$_2$ weakly bound dimers at a temperature
$T\sim 3\mu$K and $a\sim 800$\AA, the corresponding ratio is of the order of
$10^{-4}$ or $10^{-5}$. This opens wide possibilities for reaching BEC of the dimers
and cooling the Bose-condensed gas to temperatures of the order of its chemical
potential. Long-lived BEC of weakly bound dimers has been recently observed for
$^{40}$K$_{2}$ at JILA \cite{jila2,jila3} and for $^{6}$Li$_{2}$ at Innsbruck
\cite{rudy2,rudy3}, MIT \cite{mit1,mit2}, ENS \cite{ens2}, and Rice \cite{randy2} .
Measurements  of the dimer-dimer scattering length in these experiments confirm our
result $a_{dd}=0.6a$ \cite{Petrov1} with accuracy up to 30\% \cite{comment}. This
result is also confirmed by recent Monte Carlo calculations \cite{Giorgini} of the
ground state energy in the  molecular BEC regime. 

 In conclusion, we have developed a theory of elastic and inelastic collisions of
weakly bound molecules formed in a two-component atomic Fermi gas. We emphasize that
the remarkable collisional stability of these  molecules at $a\gg R_e$ is due to
{\it Fermi statistics}. This effect is not present for weakly bound molecules of
bosonic atoms, even if they have the same large size. Indeed, identical fermionic
atoms participating in the relaxation process at short interparticle distances have
very small relative momenta $k\sim 1/a$ and, hence, the process is suppressed
compared to the case of molecules of bosonic atoms.

The long lifetime of weakly bound diatomic molecules of fermionic atoms allows
interesting manipulations with these dimers. It seems realistic to arrange a deep
evaporative cooling of their Bose-condensed gas to temperatures of the order of the
chemical potential. Then, converting the molecular BEC into fermionic atoms by
adiabatically changing the scattering length to negative values, one provides an
additional cooling. The obtained atomic Fermi gas will have extremely low
temperatures $T\sim 10^{-2}E_F$ and can be already in the BCS regime \cite{carr}. At
these temperatures one has a very strong Pauli blocking of elastic collisions and
expects the collisionless regime for the thermal cloud, which is promising for
identifying the BCS-paired state through the observation of collective oscillations
or free expansion \cite{stringari}. 

Another idea is related to transferring weakly bound molecules of fermionic atoms to
their ground ro-vibrational state by using two-photon spectroscopy, as proposed in
Ref. \cite{DeM} for molecules of bosonic atoms. Long lifetime of weakly bound dimers
of fermionic atoms at densities $\sim 10^{13}$ cm$^{-3}$ should provide a much more
efficient production of ground state molecules compared to the case of 
dimers of bosonic atoms, where one has severe limitations on achievable densities and
lifetimes. 

\section*{Acknowledgments}  \label{ackn}  

We acknowledge fruitful discussions with C. Lobo. This work was supported by NSF
through a grant for the Institute for Theoretical Atomic, Molecular and Optical
Physics at Harvard University and Smithsonian Astrophysical Observatory, by the
Centre National de la Recherche Scientifique (CNRS), by the Nederlandse
Stichting voor Fundamenteel Onderzoek der Materie (FOM), by INTAS, and by the
Russian Foundation for Fundamental Research. G.V.S. acknowledges hospitality from
ITAMP where he was working on this project during his visit in 2004. This research
was also supported in part by the National Science Foundation under Grant No.
PHY99-07949 (preprint number NSF-KITP-04-88). LKB is a Unit\'e de Recherche de
l'Ecole Normale Sup\'erieure et de l'Universit\'e Paris 6, associ\'ee au CNRS. LPTMS
is a mixed research unit (No.8626) of CNRS and Universit\'e Paris Sud. 

\section*{Appendix}  \label{app}

We first prove that in the limit $r_+\rightarrow 0$ we have Eq.~(\ref{termP}) and the
integral on the right hand side of this equation is convergent. So, we consider the
integral
\begin{equation} \label{I1}
I=\int G_E(\sqrt{({\bf r}-{\bf r'})^2+r_+^2})[h({\bf r'})-h({\bf
r})]d^3r'
\end{equation}
and analyze the limiting case of $r_+\rightarrow 0$. For this purpose we use the
expansion of the 
functions $h({\bf r}')$ and $h({\bf r})$ in spherical harmonics. 
This type of expansion can be made for any function of the components of a
three-dimensional vector and it reads
\begin{equation} \label{hml}
h({\bf r})=\sum_{l=0}^{\infty}\sum_{m=-l}^{l}h_{lm}(r)Y_{lm}(\theta_{{\bf
r}},\phi_{{\bf r}}),
\end{equation}
with $\theta_{{\bf r}}$ and $\phi_{{\bf r}}$ being the polar and azimuthal angles of
the vector ${\bf r}$ 
with respect to a quantization axis. Note that in the analysis of the atom-dimer
elastic scattering  in 
Section \ref{sec.atommol} we used an expansion in Legendre polynomials for the
function $f({\bf r})$. 
The reason is that the  bound molecular state has zero orbital angular momentum and
there is a cylindrical 
symmetry in the system. For a general 3-body problem one should, in principle, use
Eq.~(\ref{hml}).  

For spherical harmonics $Y_{lm}(\theta_{{\bf r}'},\phi_{{\bf r}'})$ in the expansion
(\ref{hml}) for the function $h({\bf r}')$, 
we use the relation (see \cite{LL3}):
\begin{equation}  \label{YD}
Y_{lm}(\theta_{{\bf r}'},\phi_{{\bf
r}'})=\sum_{m'}D^{(l)}_{m'm}(\phi_{{\bf r}},\theta_{{\bf
r}},0)Y_{lm'}(\theta',\phi'),
\end{equation}
where $D^{(l)}_{m'm}$ is the matrix of finite rotations, and $\theta', \phi'$ are the
polar and  azimuthal angles of the vector ${\bf r}'$ with respect to the
quantization axis parallel to the vector ${\bf r}$. Integrating over the angle
$\phi'$ in Eq.~(\ref{I1})we  find that all terms with $m'\neq 0$ vanish.  This is
because  the
argument of the Green function $G_E$ in Eq.~(\ref{I1}) depends only on the angle
$\theta'$. 
Using the relations $Y_{l0}(\theta',0)=i^l\sqrt{(2l+1)/4\pi}P_l(\cos\theta')$ and 
$D_{0m}^{(l)}(\phi_{{\bf r}},\theta_{{\bf
r}},0)=i^{-l}\sqrt{4\pi/(2l+1)}Y_{lm}(\theta_{{\bf r}},\phi_{{\bf r}})$, 
we then reduce Eq.~(\ref{I1}) to the form
\begin{eqnarray} 
&&I=\sum_{l=0}^{\infty}\sum_{m=-l}^lI_{lm}Y_{lm}(\theta_{{\bf r}},\phi_{{\bf r}});
\label{Iex} \\
&&I_{lm}=2\pi\int_{-1}^1
d\cos{\theta'}\int_0^{\infty}r'^2dr'\Big[ h_{lm}(r')P_l(\cos{\theta'})  \nonumber \\
&&-h_{lm}(r)\Big] G_E\left(\sqrt{r^2+r'^2+r_+^2-2rr'\cos{\theta'}}\right). \label{A1} 
\end{eqnarray}
With Eq.~(\ref{general.green}) for
the Green function $G_E$, integration in parts over $d\cos{\theta'}$ transforms
Eq.~(\ref{A1}) into
\begin{eqnarray} \label{A2}
&&I_{lm}={\cal
A}_{lm}+\int_0^{\infty}\frac{\sqrt{-E}}{4\pi^2}\frac{r'dr'}{r}[h_{lm}(r')-h_{lm}(r)]
\nonumber \\
&&\frac{K_1\left(\sqrt{-E}\sqrt{(r-r')^2+r_+^2}\right)}{\sqrt{(r-r')^2+r_+^2}}.
 \end{eqnarray}
The quantities ${\cal A}_{lm}$ represent the sum of integrals in which for $r_+=0$
and
$r'\rightarrow r$ the integrand either remains finite or contains a logarithmic
integrable singularity. Setting $r_+=0$ one obtains finite values of these
quantities:
\begin{align} \label{Acal}
&{\cal A}_{lm}=\int_0^{\infty}\frac{r'}{r}dr'\frac
{[h_{lm}(r)+(-1)^{l+1}h_{lm}(r')]\sqrt{-E}}{4\pi^2(r+r')} \nonumber \\
&\times K_1\left(\sqrt{-E}(r+r')\right)
+\int_0^{\infty}\frac{h_{lm}(r')}{4\pi^2}\frac{dr'}{r^2}\Big(
\frac{l(l+1)}{2} \nonumber \\
&\times\left[K_0\left(\sqrt{-E}|r-r'|\right)+(-1)^lK_0\left(\sqrt{-E}(r+r')\right)\right]
\nonumber \\
&-\int_{-1}^1dxP_l^{''}(x)K_0\left(\sqrt{-E}\sqrt{r^2+r'^2-2rr'x}\right)\Big).
\end{align}
Equations (\ref{A2}) and (\ref{Acal}) are written for the case $E<0$. For $E>0$ one
should put in these equations $E$ instead of $-E$ and replace the decaying Bessel 
functions $K_1$ and $K_0$ by the Hankel functions $2iH_1$ and $2iH_0$, respectively. 

For calculating the integral in Eq.~(\ref{A2}) we divide the region of integration
into two parts: inside a small interval ${\cal L}$, where
$|r'-r|\ll 1/\sqrt{|E|}$, and outside this interval. In addition, we require that the
function $h_{lm}(r')$ does not significantly change inside the interval ${\cal L}$.
The
integral outside the interval ${\cal L}$ remains finite and independent of $r_+$ in
the
limit $r_+\rightarrow 0$. Inside the interval ${\cal L}$ the argument of the Bessel
(or Hankel) function is
small and the integral in Eq.~(\ref{A2}) reduces to
$$\frac{1}{4\pi^2}\int_{r'\in{\cal
L}}\frac{[h_{lm}(r')-h_{lm}(r)]}{(r'-r)^2+r_+^2}\frac{r'}{r}dr'.$$
We then expand the quantity $[h_{lm}(r')-h_{lm}(r)]$ in powers of $(r'-r)$. Quadratic
and
higher order terms of the expansion give contributions to the integrand, which
remain finite for $r_+\rightarrow 0$. The contribution of the linear term is given
by 
\begin{equation} \label{A3}
\frac{1}{4\pi^2}\frac{dh_{lm}(r)}{dr}\int_{r'\in{\cal L}}\frac{r'-r}
{(r'-r)^2+r_+^2}\frac{r'}{r}dr'.
\end{equation}
For $r_+\rightarrow 0$ the integral in Eq.~(\ref{A3}) remains finite and can be
written as
$$\int_{r'\in{\cal L}}\frac{dr'}{r}+P\int_{r'\in{\cal L}}\frac{dr'}{(r'-r)},$$
where the symbol $P$ denotes the principal value of the integral. We thus see that in
the limit $r_+\rightarrow 0$ the whole integral $I$ given by Eq.~(\ref{I1}) is
finite. Setting $r_+=0$ in the initial expression (\ref{I1}) and keeping the symbol
$P$ for the integration over $dr'$, one then writes the integral $I$ in the form of
Eq.~(\ref{termP}):
$$I=P\int G_E(|{\bf r}-{\bf r}'|)[h({\bf r'})-h({\bf r})]d^3r'.$$

We now derive Eq.~(\ref{reg2}) for the 4-body problem of the dimer-dimer scattering 
and prove that the  integral in the second line of
this equation is convergent. So, changing the notation from ${\bf r}_2$ to ${\bf r}$,
we start with the integral 
\begin{equation} \label{J}
J=\int d^3r'd^3R'[h({\bf r}',{\bf R}')-h({\bf r},{\bf R})]G(|S-S_1|),
\end{equation}
where the argument  of the Green function in Eq.~(\ref{J}) is  
$$|S-S_1|=\sqrt{r_1^2+({\bf r}-{\bf r}')^2+({\bf R}-{\bf R}')^2},$$
and consider the limit $r_1\rightarrow 0$. As in the 3-body case, we expand the
function $h({\bf r},{\bf R})$ in spherical harmonics. The expansion now reads:
\begin{eqnarray}  \label{hrR}
&&h({\bf r},{\bf R})=\sum_{l=0,l'=0}^{\infty}\sum_{m=-l}^{l}\sum_{m'=-l'}^{l'}
h_{lm}^{l'm'}(r,R)  \nonumber \\
&&\times Y_{lm}(\theta_{{\bf r}},\phi_{{\bf r}})Y_{l'm'}(\theta_{{\bf R}},
\phi_{{\bf R}}),
\end{eqnarray}
where $\theta_{{\bf r}},\phi_{{\bf r}}$ and $\theta_{{\bf R}},\phi_{{\bf R}}$ are
the polar and azimuthal angles of the vectors ${\bf r}$ and ${\bf R}$. Then, using
Eq.~(\ref{YD}) for each of the spherical harmonics in the expansion of the function
$h({\bf r}',{\bf R}')$, we transform Eq.~(\ref{J}) to the form
\begin{align}
&J=\sum_{l=0,l'=0}^{\infty}\sum_{m=-l}^{l}\sum_{m'=-l'}^{l'}J_{lm}^{l'm'}
Y_{lm}(\theta_{{\bf r}},\phi_{{\bf r}})Y_{l'm'}(\theta_{{\bf R}},
\phi_{{\bf R}}); \label{Jexpanded} \\
&J_{lm}^{l'm'}=4\pi^2\int_{-1}^1d\cos\theta'_{{\bf r}}\int_{-1}^1d\cos\theta'_{{\bf
R}}
\int_0^{\infty}r'^2dr'\int_0^{\infty}\!\!R'^2dR' \nonumber \\
&\times [h_{lm}^{l'm'}(r',R')P_l(\cos\theta'_{{\bf r}})P_{l'}(\cos\theta'_{{\bf
R}})-h_{lm}^{l'm'}(r,R)]G(X),  \label{Jlm}
\end{align}  
with the angle between the vectors ${\bf r}'$ and ${\bf r}$ denoted as $\theta'_{{\bf
r}}$, 
the angle between ${\bf R}'$ and ${\bf R}$ as $\theta'_{{\bf R}}$, and
the argument of the Green function written as
$$X\!=\!\sqrt{r_1^2\!+\!r^2\!+\!r'^2\!-\!2rr'\cos\theta'_{{\bf
r}}\!+\!R^2\!+\!R'^2\!-\!2RR'\cos\theta'_{{\bf R}}}
.$$

We then separate out the region of integration $\tilde V$, 
where the angles $\theta'_{{\bf r}}$ and $\theta'_{{\bf R}}$
are small, $r'$ is close to $r$, and $R'$ is close to $R$. The region $\tilde V$ is
determined
by the relations:
\begin{eqnarray}
&&|r'-r|\leq r_0\ll r',r,a;  \label{rr} \\
&&|R'-R|\leq r_0\ll R',R,a; \label{RR} \\
&&\theta'_{{\bf r}}\leq\eta_{{\bf r}}\ll {\rm min}\{1,a/\sqrt{rr'}\}; \label{thetar}
\\
&&\theta'_{{\bf R}}\leq\eta_{{\bf R}}\ll {\rm min}\{1,a/\sqrt{RR'}\}, \label{thetaR}
\end{eqnarray}
and the quantities $\eta_{{\bf r}},\eta_{{\bf R}}$ are selected such that they
satisfy the inequalities
\begin{eqnarray}
&&\eta_{{\bf r}}\sqrt{rr'}\gg |r'-r|, \label{ineqr}\\
&&\eta_{{\bf R}}\sqrt{RR'}\gg |R'-R|. \label{ineqR}
\end{eqnarray}
Outside the region $\tilde V$ one can directly put $r_1=0$ in the argument of the
Green
function and see that the quantities $J_{lm}^{l'm'}$ remain finite. These
contributions are denoted
below as ${\bar J}_{lm}^{l'm'}$. In the region $\tilde V$ the argument $X$ of the
Green 
function can become equal to $r_1$ and, in principle, this region can be thought of
as 
the one leading to a singular behavior of the integral $J$ in the limit
$r_1\rightarrow 0$.
We will show that this not the case.

The Green function $G$ is given by Eq.~(\ref{Green4body}), and in the region $\tilde
V$ the argument of the 
decaying Bessel function, $\sqrt{2}X/a$, is small. Then, for small angles
$\theta_{{\bf r}}$ and
$\theta_{{\bf R}}$, from Eq.~(\ref{Green4body}) we obtain
\begin{equation}  \label{G4small}
\!G\!=\!\frac{15}{32\pi^4}(r_1^2\!+\!(r'\!-\!r)^2\!+\!(R'\!-\!R)^2\!+\!\theta^{'2}_{{\bf
r}}rr'
\!+\!\theta^{'2}_{{\bf R}}RR')^{-7/2}.
\end{equation}    
Inside the region $\tilde V$ the inequalities (\ref{thetar}) and (\ref{thetaR}) allow
us to put 
$P_l(\cos\theta'_{{\bf r}})=P_{l'}(\cos\theta'_{{\bf R}})=1$ in Eq.~(\ref{Jlm}). On
the other hand, 
owing to the inequalities (\ref{ineqr}) and (\ref{ineqR}), we can put infinity for
the upper limits of integration
over $\theta'_{{\bf r}}$ and $\theta'_{{\bf R}}$. Then, integrating over the angles
inside the region $\tilde V$ and 
taking into account the contribution ${\bar J}_{lm}^{l'm'}$ from the configuration
space outside this region,
we reduce Eq.~(\ref{Jlm}) to
\begin{eqnarray} \label{Jlmreduced}
&&J_{lm}^{l'm'}={\bar J}_{lm}^{l'm'}+\frac{1}{8\pi^2}\int_{r',R'\in\tilde
V}\frac{r'R'}{rR}
dr'dR'  \nonumber \\
&&\times\frac{[h_{lm}^{l'm'}(r',R')-h_{lm}^{l'm'}(r,R)]}{(r_1^2+(r'-r)^2+(R'-R)^2)^{3/2}}.
\end{eqnarray} 

For calculating the integral in Eq.~(\ref{Jlmreduced}) we expand the function
$h(r',R')$ in powers of
$(r'-r)$ and $(R'-R)$. Quadratic and higher order terms of the expansion lead to the
integrand which 
remains finite for $r_1\rightarrow 0$, at least after a straightforward integration
over one of the
variables, $r'$ or $R'$. For example, setting $r_1=0$ quadratic terms yield
\begin{align} 
&\frac{\partial^2h_{lm}^{l'm'}(r,R)}{8\pi^2\partial r^2}\int_{r'\in \tilde V}
\frac{r_0dr'}{(r_0^2+(r'-r)^2)^{1/2}} \nonumber \\
&+\frac{\partial^2h_{lm}^{l'm'}(r,R)}{8\pi^2\partial R^2}\int_{R'\in \tilde
V}\frac{r_0dR'}{(r_0^2+(R'-R)^2)^{1/2}}
\nonumber \\
&+\frac{\partial^2h_{lm}^{l'm'}(r,R)}{4\pi^2\partial r\partial R}\int_{r'\in\tilde
V}\frac{dr'(r'-r)^2}{rR} \nonumber \\
&\times\left[\ln\left(\frac{r_0+\sqrt{r_0^2+(r'-r)^2}}{|r'-r|}\right) 
-\frac{r_0}{(r_0^2+(r'-r)^2)^{1/2}}\right]. \nonumber
\end{align}

The contribution of the linear terms of the expansion, after integrating over $R'$, 
can be written as  
\begin{align}  \label{linearterms}
&\frac{\partial h_{lm}^{l'm'}(r,R)}{4\pi^2r\partial r}\int_{r'\in\tilde
V}\frac{(r'-r)}{r_1^2+(r'-r)^2}\frac{r_0r'dr'}
{(r_0^2+(r'-r)^2)^{1/2}} \nonumber \\
&+\frac{\partial h_{lm}^{l'm'}(r,R)}{4\pi^2R\partial R}\int_{r'\in\tilde
V}\frac{r'dr'}{r}
\Big[\ln\left(\frac{r_0+\sqrt{r_0^2+(r'-r)^2}}{|r'-r|}\right) \nonumber \\
&-\frac{r_0}{(r_0^2+(r'-r)^2)^{1/2}}\Big],
\end{align}
where we put $r_1=0$ in the second integral as it then contains only an integrable
logarithmic singularity
for $r'\rightarrow r$. The first integral on the right hand side of
Eq.~(\ref{linearterms}) also remains finite
in the limit of $r_1\rightarrow 0$. It can be written in the form:
\begin{eqnarray}
&&\int_{r'\in\tilde V}\frac{r_0dr'}{(r_0^2+(r'-r)^2)^{1/2}} \nonumber \\
&&+P\int_{r'\in\tilde V}\frac{rdr'}{(r'-r)}\,\frac{r_0}{(r_0^2+(r'-r)^2)^{1/2}},
\nonumber
\end{eqnarray}
with the symbol $P$ denoting the principal value of the integral. This means that
setting $r_1=0$,
each of the quantities $J_{lm}^{l'm'}$ is equal to the principal value of the
integral in 
Eq.~(\ref{Jlm}), taken with respect to the integration over $dr'$. Note that we could
first make the integration over 
$dr'$ and reduce the result to the principal value for the integration over $dR'$. 
One thus sees that for $r_1\rightarrow 0$ the initial integral $J$ (\ref{J})  
can be represented in the form of Eq.~(\ref{reg2}):
\begin{eqnarray}
J=P\int d^3r'd^3R'[h({\bf r}',{\bf R}')-h({\bf r},{\bf R})]G(|{\bar S}-S_1|),
\nonumber
\end{eqnarray}      
where the symbol $P$ denotes the principal value for the integration over $dr'$ (or
$dR'$), and 
$|{\bar S}-S_1|=\sqrt{({\bf r}-{\bf r}')^2+({\bf R}-{\bf R}')^2}$.


\begin{thebibliography}{99}

\bibitem{Eag} D.M. Eagles, Phys. Rev. {\bf 186}, 456 (1969).

\bibitem{Leg} A.J. Leggett, in {\it Modern Trends in the Theory of Condensed
Matter}, edited by A. Pekalski and J. Przystawa (Springer, Berlin,1980).

\bibitem{Noz} P. Nozieres and S. Schmitt-Rink, J. Low Temp. Phys. {\bf 59},
195 (1985).

\bibitem{Rand} See for review M. Randeria, in {\it Bose-Einstein
Condensation}, edited by A. Griffin, D.W. Snoke, and S. Stringari (Cambridge
University Press, Cambridge,1995).

\bibitem{M} K. Miyake, Progr. Theor. Phys. {\bf 69}, 1794 (1983).

\bibitem{MYu} See for review M.Yu. Kagan, Sov. Physics Uspekhi {\bf 37}, 69
(1994).

\bibitem{Holland1} M. Holland, S.J.J.M.F. Kokkelmans, M.L. Chiofalo, and R. Walser,
Phys. Rev. Lett. {\bf 87}, 120406 (2001).

\bibitem{Tim} E. Timmermans, K. Furuya, P.W. Milonni, A.K. Kerman, Phys. Lett. A {\bf
285}, 228 (2001).

\bibitem{Holland2} M.L. Chiofalo, S.J.J.M.F. Kokkelmans, J.N. Milstein, and M.J.
Holland, Phys. Rev. Lett. {\bf 88}, 090402 (2002); S.J.J.M.F. Kokkelmans, J.N.
Milstein, M.L. Chiofalo, R. Walser, and M.J. Holland, Phys. Rev. A {\bf 65}, 053617
(2002).

\bibitem{Holland3} J.N. Milstein, S.J.J.M.F. Kokkelmans, and M.J. Holland, Phys. Rev.
A {\bf 66}, 043604 (2002).

\bibitem{Griffin} Y. Ohashi and A. Griffin, Phys. Rev. Lett. {\bf 89}, 130402
(2002);  Phys. Rev. A {\bf 67}, 033603 (2003).

\bibitem{Heis} H. Heiselberg, Phys. Rev. A {\bf 63}, 043606 (2001).

\bibitem{Carlson} J. Carlson, S.-Y. Chang, V.R. Pandharipande, and K.E. Schmidt,
Phys. Rev. Lett. {\bf 91}, 050401 (2003). 

\bibitem{duke1} K.M. O'Hara, S.L. Hemmer, M.E. Gehm, S.R. Granade, and J.E. Thomas,
Science {\bf 298}, 2179 (2002); M.E. Gehm, S.L. Hemmer, S.R. Granade, K.M. O'Hara,
and J.E. Thomas, Phys. Rev. A {\bf 68} 011401 (2003).

\bibitem{Bourdel} T. Bourdel, J. Cubizolles, L. Khaykovich, K.M.F. Magalh\~aes,
S.J.J.M.F. Kokkelmans, G.V. Shlyapnikov, and C. Salomon, Phys. Rev. Lett. {\bf 91},
020402 (2003).

\bibitem{Ho1} Tin-Lun Ho, Phys. Rev. Lett. {\bf 92}, 090402 (2004).

\bibitem{Falco} G.M. Falco and H.T.C. Stoof, Phys. Rev. Lett. {\bf 92}, 130401
(2004).

\bibitem{Bruun} G.M. Bruun and C. Pethik, Phys. Rev. Lett. {\bf 92}, 140404 (2004).

\bibitem{Bulgac} A. Bulgac, cond-mat/0309358.

\bibitem{Perali} A. Perali, P. Pieri, L. Pisani, G.C. Strinati, Phys. Rev. Lett. {\bf
92}, 220404 (2004).

\bibitem{Chen} Q. Chen, J. Stajic, S. Tan, and K. Levin, cond-mat/0404274.

\bibitem{Holland4} M.J. Holland, C. Menotti, and L. Viverit, cond-mat/0404234.

\bibitem{Ho2} R. Diener and Tin-Lun Ho, cond-mat/0404517.

\bibitem{ens1} J. Cubizolles, T. Bourdel, S.J.J.M.F. Kokkelmans, G.V. Shlyapnikov,
and C. Salomon, Phys. Rev. Lett. {\bf 91}, 240401 (2003).

\bibitem{rudy1} S. Jochim, M. Bartenstein, A. Altmeyer, G. Hendl, C. Chin, J.H.
Denschlag, and R. Grimm, Phys. Rev. Lett. {\bf 91}, 240402 (2003).

\bibitem{randy1} K.E. Strecker, G.B. Partridge, and R.G. Hulet, Phys. Rev. Lett. {\bf
91}, 080406 (2003).

\bibitem{jila1} C.A. Regal, M. Greiner, and D.S. Jin, Phys. Rev. Lett. {\bf 92},
083201 (2004).

\bibitem{jila2} M. Greiner, C. Regal, and D.S. Jin, Nature {\bf 426}, 537 (2003).

\bibitem{rudy2} S. Jochim, M. Bartenstein, A. Altmeyer, G. Hendl, S. Riedl, C. Chin,
J.H. Denschlag, and R. Grimm, Science {\bf 302}, 2101 (2003); M. Bartenstein, A.
Altmeyer, S. Riedl, S. Jochim, C. Chin, J.H. Denschlag, and R. Grimm, Phys. Rev.
Lett. {\bf 92}, 120401 (2004).

\bibitem{mit1} M.W. Zwierlein, C.A. Stan, C.H. Schunck, S.M.F. Raupach, S. Gupta, Z.
Hadzibabic, and W. Ketterle, Phys. Rev. Lett. {\bf 91}, 250401 (2003).

\bibitem{ens2} T. Bourdel, L. Khaykovich, J. Cubizolles, J. Zhang, F. Chevy, M.
Teichmann, L. Tarruell, S.J.J.M.F. Kokkelmans, and C. Salomon, cond-mat/0403091.

\bibitem{randy2} R.G. Hulet, KITP Conference on Quantum Gases, Santa Barbara, May
10-14, 2004.

\bibitem{jila3} C.A. Regal, M. Greiner, and D.S. Jin, Phys. Rev. Lett. {\bf 92},
040403 (2004).

\bibitem{mit2} M.W. Zwerlein, C.A. Stan, C.H. Schunck, S.M.F. Raupach, A.J. Kerman,
and W. Ketterle, Phys. Rev. Lett. {\bf 92}, 120403 (2004).

\bibitem{duke2} J. Kinast, S.L. Hemmer, M.E. Gehm, A. Turlapov, and J.E. Thomas,
Phys. Rev. Lett. {\bf 92}, 150402 (2004).

\bibitem{rudy3} M. Bartenstein, A. Altmeyer, S. Riedl, S. Jochim, C. Chin, J.H.
Denschlag, and R. Grimm, Phys. Rev. Lett. {\bf 92}, 203201 (2004).

\bibitem{rudy4} C. Chin, M. Bartenstein, A. Altmeyer, S. Riedl, S. Jochim, J.H.
Denschlag, and R. Grimm, cond-mat/0405632.

\bibitem{Petrov1} D.S. Petrov, C. Salomon, and G.V. Shlyapnikov, cond-mat/0309010.

\bibitem{Bethe} H. Bethe and R. Peierls, Proc. Roy. Soc. (London) {\bf 148A}, 146
(1935). 

\bibitem{Huang} K. Huang, {\it Statistical Mechanics} (Wiley, New York, 1987).

\bibitem{LL3} L.D. Landau and E.M. Lifshitz, {\it Quantum Mechanics},
(Butterworth-Heinemann, Oxford, 1999).

\bibitem{fedorov} E. Nielsen, D.V. Fedorov, and A.S. Jensen, Phys. Rep. {\bf 347},
374 (2000).

\bibitem{efimov} V.N. Efimov, Yad. Fiz. {\bf 12}, 1080 (1970) [Sov. J. Nucl. Phys.
{\bf 12}, 589 (1971)]; Nucl. Phys. A {\bf 210}, 157 (1973).

\bibitem{STM} G.V. Skorniakov and K.A. Ter-Martirosian, Zh. Eksp. Teor. Phys. {\bf
31}, 775 (1956) [Sov. Phys. JETP {\bf 4}, 648 (1957)].

\bibitem{Danilov} G.S. Danilov, Sov. Phys. JETP {\bf 13}, 349 (1961).

\bibitem{Petrov2} D.S. Petrov, Phys. Rev. A {\bf 67}, 010703 (2003).

\bibitem{strinati} P. Pieri and G.C. Strinati, Phys. Rev. B {\bf 61}, 15370 (2000).

\bibitem{Breit} G. Breit and E. Wigner, Phys. Rev. {\bf 49}, 519 (1936).

\bibitem{Feshbach} H. Feshbach, Ann. Phys. {\bf 19}, 287 (1962); {\it Theoretical
Nuclear Physics} (Wiley, New York, 1992).

\bibitem{Fano} U. Fano, Phys. Rev. {\bf 124}, 1866 (1961).

\bibitem{Verhaar} A.J. Moerdijk, B.J. Verhaar, and A. Axelsson, Phys. Rev. A {\bf
51}, 4852 (1995); see also E. Tiesinga, B.J. Verhaar, and H.T.C. Stoof, Phys. Rev. A
{\bf 47}, 4114 (1993).

\bibitem{Gora} P.O. Fedichev. Yu. Kagan, G.V. Shlyapnikov, and J.T.M. Walraven, Phys.
Rev. Lett. {\bf 77}, 2913 (1996).

\bibitem{Bohn} J.L. Bohn and P.S. Julienne, Phys. Rev. A {\bf 60}, 414 (1999).

\bibitem{Bruun2} G.M. Bruun, cond-mat/0401497.

\bibitem{Petrovbosons} D.S. Petrov, cond-mat/0404036.

\bibitem{Palo} S. De Palo, M.L. Chiofalo, M.J. Holland, and S.J.J.M.F. Kokkelmans,
cond-mat/0404672.

\bibitem{Eric} E.A. Cornell, KITP Conference on Quantum Gases, Santa Barbara, May
10-14, 2004.

\bibitem{Ho3} R. Diener and Tin-Lun Ho, cond-mat/0405174.

\bibitem{Servaas} S.J.J.M.F. Kokkelmans, G.V. Shlyapnikov, and C. Salomon, Phys. Rev.
A {\bf 69} 031602 (2004). 

\bibitem{comment} The revised version of cond-mat/0306302 by A.Bulgac, P.F. Bedaque,
and A.C. Fonseca, which appeared on February 16 of 2004, contains a statement that
these authors numerically found the result  $a_{dd}=0.6a$ derived in our work
\cite{Petrov1}. 

\bibitem{Giorgini} G.E. Astrakharchik, J. Boronat, J. Casulleras, S. Giorgini,
cond-mat/0406113.

\bibitem{carr} L.D. Carr, G.V. Shlyapnikov, and Y. Castin, Phys. Rev. Lett. {\bf 92},
150404 (2004).

\bibitem{stringari} C. Menotti, P. Pedri, and S. Stringari, Phys. Rev. Lett. {\bf
89}, 250402 (2002).

\bibitem{DeM} A.J. Kerman, J.M. Sage, S. Sainis, T. Bergeman, and D. DeMille, Phys.
Rev. Lett. {\bf 92}, 153001 (2004).



\end{thebibliography}
\end{document}